\documentclass[a4paper,11pt]{article}
\pdfoutput=1
\usepackage{color}
\usepackage{graphicx}
\usepackage{epstopdf}
\usepackage{amsmath}
\usepackage{ulem}
\usepackage[labelsep=quad, font={small}, labelfont=bf, width=0.8\textwidth, aboveskip=0.4cm, belowskip=0cm]{caption}
\usepackage[bottom = 3cm, top = 3cm, left = 3cm, right = 3cm, headheight = 14pt]{geometry}
\usepackage{multirow}
\usepackage{array}
\usepackage{fancyhdr}
\usepackage{upgreek}
\usepackage{rotating}
\usepackage{booktabs}
\usepackage{lscape}
\usepackage{subcaption}

\newcommand{\ket}[1]{| #1 \rangle}

\begin{document}

\title{All-fibre multiplexed source of high-purity \\ heralded single photons}
\author{Robert J. A. Francis-Jones, Rowan A. Hoggarth, and Peter J. Mosley \\ Centre for Photonics and Photonics Materials, Department of Physics, \\ University of Bath, Bath BA2 7AY, UK \\ R.J.A.Francis-Jones@bath.ac.uk}
%\date{}

\maketitle
\thispagestyle{empty}

\noindent

\textbf{Single photon sources based on spontaneous photon-pair generation have enabled pioneering experiments in quantum optics~\cite{Pan2012Multiphoton_Entanglement}. However, their non-determinism presents a bottleneck to scaling up photonic and hybrid quantum-enhanced technologies~\cite{Ladd2010Quantum_Computers}. Furthermore, photon pairs are typically emitted into many correlated frequency modes, producing an undesirable mixed state on heralding~\cite{Grice2001Eliminating-Frequency-and-Space-Time}. Here we present a complete fibre-integrated heralded single photon source that addresses both these difficulties simultaneously. We use active switching to provide a path to deterministic operation by multiplexing separate spontaneous sources, and dispersion engineering to minimise frequency correlation for high-purity single photon generation. All the essential elements -- nonlinear material with dispersion control, wavelength isolation, optical delay, and fast switching -- are incorporated in a low-loss alignment-free package that heralds photons in telecoms single-mode fibre. Our results demonstrate a scalable approach to delivering pure single photons directly into guided-wave photonic devices.}

%\cite{Somaschi2015Near-optimal-single} \cite{Ding2016On-Demand-Single-Photons}

Single photons are a vital resource for quantum-enhanced technologies, not only those that exploit nonclassical light to make high-precision measurements or process quantum information but also hybrid platforms that require photonic links between matter qubits. Spontaneous photon-pair sources based on either parametric downconversion (PDC) or four-wave mixing (FWM) combined with heralding detection have established themselves as the workhorses of nonclassical light generation due to their high performance and the relative simplicity of the apparatus required~\cite{Burnham1970Observation-of-Simultaneity,Li2004All-Fiber-Photon-Pair-Sources}: a laser, typically pulsed, generates signal and idler photon pairs as it propagates through a nonlinear medium. Whether or not their ultimate performance exceeds that of single-emitter photon sources~\cite{Santori2002Indistinguishable_Photons_From,Englund2010Deterministic_Coupling_of_a,Nilsson2013Quantum_Teleportation_Using}, photon-pair sources are essential to provide capability for the medium-term development of quantum technologies, and perhaps further into the future. However, we have reached the limits of the intrinsic capabilities of these sources; it is therefore of paramount importance that we make every effort to enhance their performance.

Two limitations arise from both pair generation mechanisms. Firstly, the number of photon pairs produced per mode follows a thermal distribution. The resulting probability $p_1$ of generating one and only one photon pair from a single pump laser pulse can never be higher than $p_1 = 0.25$~\cite{Christ2012Limits-on-the-Deterministic}. Typically, the probability is kept much lower ($p_1 \approx 0.01$) to limit detrimental contributions from $n>1$ multi-pair events which increase of order $p_1^n$. The resulting low single photon delivery probability, $\eta_h \eta_s p_1$ where $\eta_h$ is the lumped heralding detector channel efficiency and $\eta_s$ is the signal channel transmission, is a serious difficulty for larger-scale devices that require the simultaneous delivery of several heralded single photons from independent sources~\cite{Yao2012Observation-of-Eight-Photon}. The overall probability of delivering $N$ independent single photons, $(\eta_h \eta_s p_1)^N$, falls rapidly as the number of sources $N$ increases. In order to alleviate this problem, it is necessary to break the link between the probability of delivering a heralded single photon and that of generating multiple pairs. One solution is a high-efficiency quantum memory with controllable read-out to synchronise photons produced independently during different clock cycles~\cite{Nunn2013Enhancing-Multiphoton-Rates}. Alternatively a multiplexed source can be built to route the heralded output of several modes into a single mode conditioned on a heralding detection~\cite{Migdall2002Tailoring-Single-Photon}. Both methods enable the probability of delivering a single photon to be increased without a commensurate increase in the multi-photon probability~\cite{Adam2014Optimisation_of_Periodic,Bonneau2014Effect_of_loss_on,Francis-Jones2014Exploting_The_Limits}, though multiplexing is less technologically challenging as it requires only low-loss optical delay and active switching combined with fast detection and feed-forward~\cite{Ma2011Experimental_Generation_of,Collins2013Integrated_Spatial_Multiplexing,Meany2014Hybrid-photonic-circuit,Kaneda2015Time_Multiplexed_Heralded,Mendoza2016Active_Tempora_and_Spatial}. It has been shown that multiplexing independent sources can yield significant performance enhancements, and even approach quasi-deterministic operation \cite{Christ2012Limits-on-the-Deterministic}.

The second limitation originates from energy conservation between the pump and daughter fields. The redistribution of energy from pump to daughter photons typically produces anti-correlation between the signal and idler frequencies. A heralding detection of one photon then provides information about its twin, and the frequency mode into which the heralded photon is projected varies randomly from shot to shot. The resulting mixed state can be cleaned up with narrow-band spectral filtering, but at the cost of significant loss and with a purity that approaches unity only as the filter bandwidth tends to zero~\cite{Silverstone2014On-Chip-Quantum-Interference}. More elegantly, by tailoring the dispersion of the nonlinear medium and using ultrashort pump pulses, frequency correlation can be minimised at the point of generation yielding a high-purity state directly upon heralding~\cite{Mosley2008Heralded-Generation-of-Ultrafast,Halder2009Nonclassical_Two_Photon,Cohen2009Tailored-Photon-Pair-Generation,Soller2011High-Performance-Single-Photon,Eckstein2011Highly_Efficient_Single_Pass}. By expressing the joint spectral amplitude of the photon pair, $f(\omega_s,\omega_i)$, in the Schmidt basis of frequency modes $\xi_s(\omega_s)$ and $\zeta_i(\omega_i)$:
\begin{equation}
\ket{\Psi} = \int\int d\omega_i d\omega_s f(\omega_s,\omega_i) \ket{\omega_s}\ket{\omega_i} \rightarrow f(\omega_s,\omega_i) = \sum_n \sqrt{\lambda_j} \xi_s(\omega_s) \zeta_i(\omega_i),
\end{equation}
the level of frequency correlation, and hence the reduced-state purity $P$, can be found from the cooperativity parameter $K = 1/\sum_j \lambda_j^2 = 1/P$. A completely separable state will have only one pair of Schmidt modes active such that $K = P = 1$, but as the number of Schmidt modes increases $K \rightarrow \infty$ and $P \rightarrow 0$.

% discuss group-velocity matching? Or in supp info?

\begin{figure*}
	\centering
	\includegraphics[width = 16.5cm]{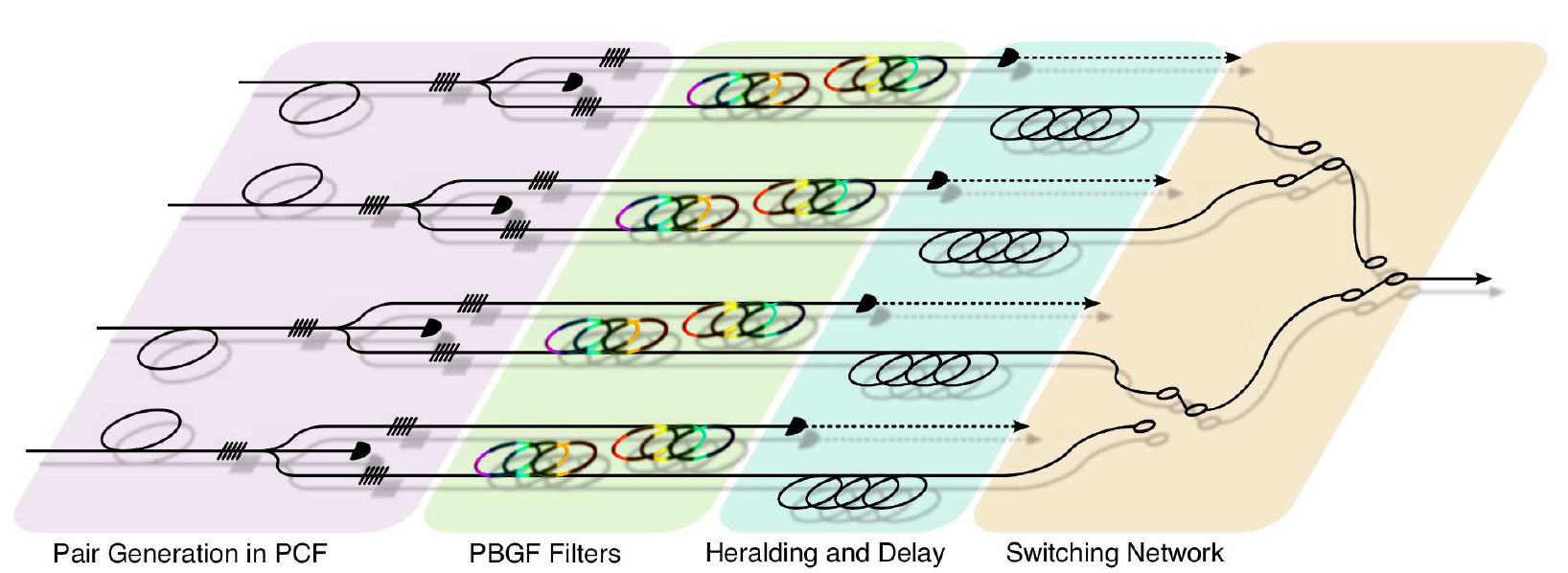}%{General_Schematic/Diagram_wo_boxes_w_shadow6}
    \caption{Schematic depicting the spatial multiplexing of four heralded single photon sources using a network of 2$\times$1 switches.}
    \label{fig:Multi_Source_Schematic}
\end{figure*}

Our heralded single-photon source addresses both of these limitations simultaneously in a fully-integrated fibre architecture. In order to multiplex single photons efficiently, we require high-purity photons to be heralded in modes that can be delayed in low-loss telecoms fibre. We have developed sources that achieve this through the application of bespoke microstructured fibre technology: using dispersion-engineered photonic crystal fibre (PCF) and novel broadband photonic bandgap fibre (PBGF) filters for wavelength isolation we have generated heralded photons in a fully guided-wave system with a high level of intrinsic purity. With a low-loss fibre-pigtailed switch we multiplexed the output of two sources to enhance the single-photon delivery probability over that of a single source without any increase in multi-photon noise.

% Mathematical intro - purity, visibility, multiplexing simple case

%\subsection*{High-purity photons in fibre}

\begin{figure*}
	\centering
        \begin{subfigure}[b]{4cm}
    	\centering
        \includegraphics[width = \textwidth]{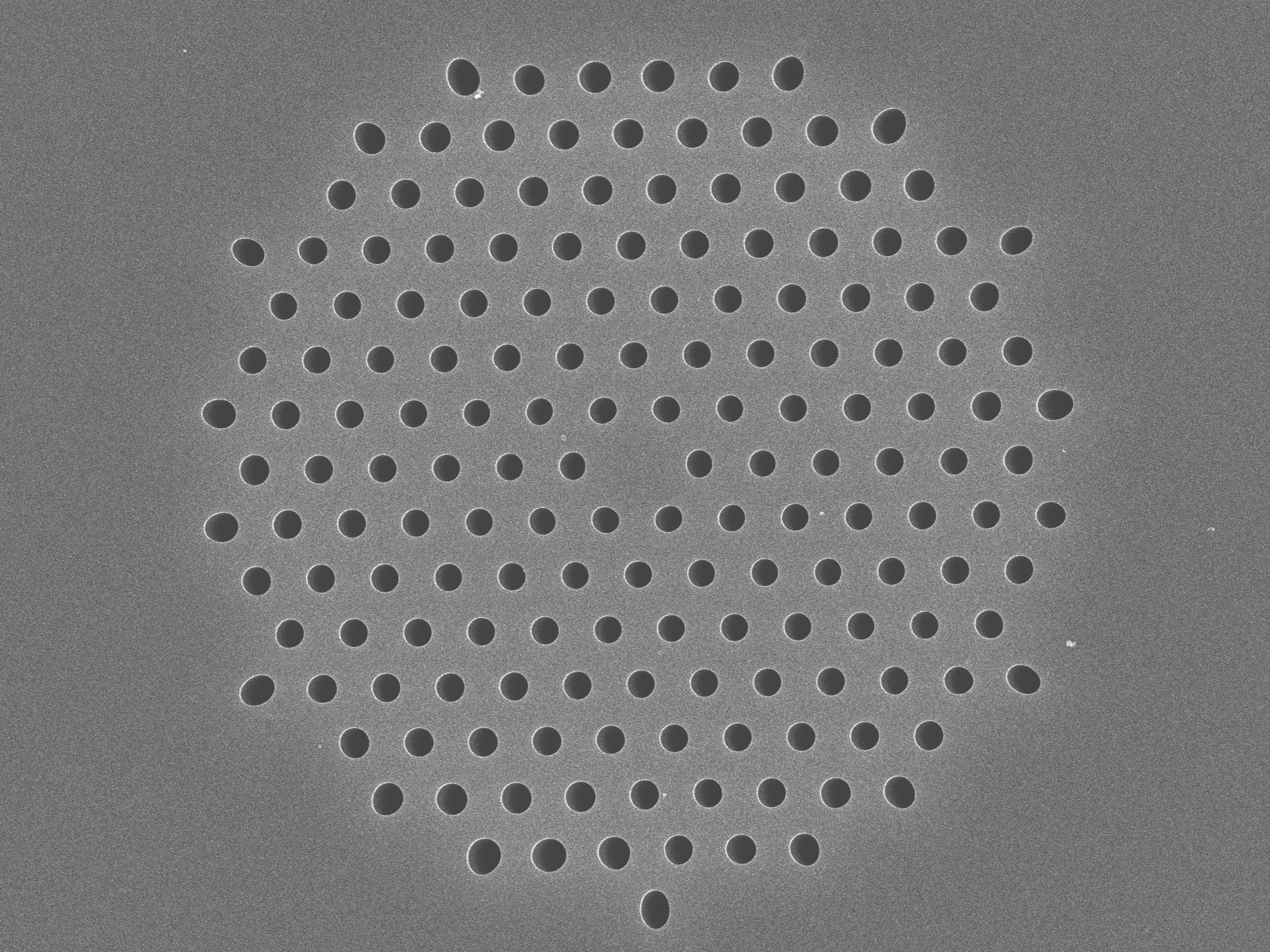}%{Band7_PCF_SEM_small}
        \caption{}
        \label{fig:PCF_SEM}
   \end{subfigure} \hspace{5mm}
  \begin{subfigure}[b]{9cm}
    	\centering
        \includegraphics[width = \textwidth]{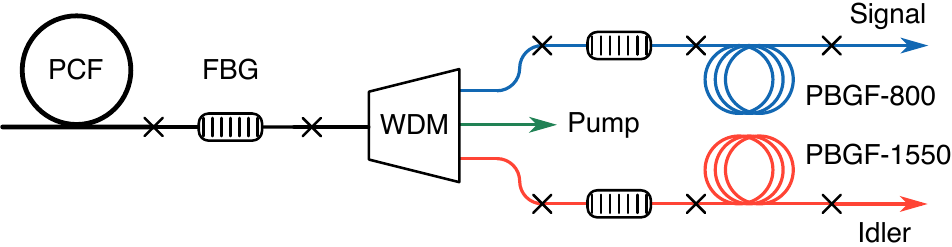}%{Single_Source_Schematic}
        \label{fig:Single_Source_Schematic}
        \caption{}
   \end{subfigure}

   \begin{subfigure}[b]{8cm}
      	\centering
      	\includegraphics[width = \textwidth]{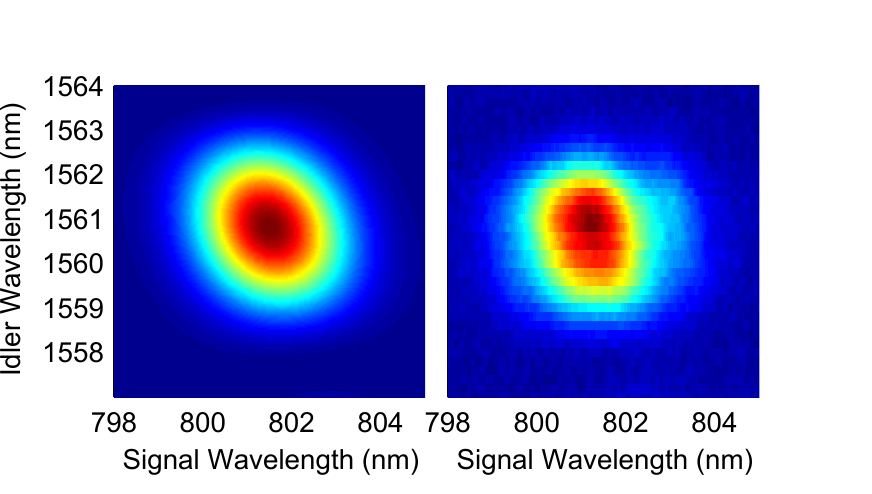}%{JSI/JSI_Stim_Paired_9x5}%JSI_Sim_Paper_nm_f_35_5x5}
      	\caption{}
      	\label{fig:JSI_Upper}
   	\end{subfigure}
%   	\begin{subfigure}[b]{5cm}
%   			\centering
%   		 	\includegraphics[width = 5cm]{JSI/JSI_Stim_S1_Paper_F_35_5x5_no_y}
%    			\caption{}
%   			\label{fig:JSI_Sim}
%   \end{subfigure}
    \begin{subfigure}[b]{6cm}
   		\centering
        \includegraphics[width = \textwidth]{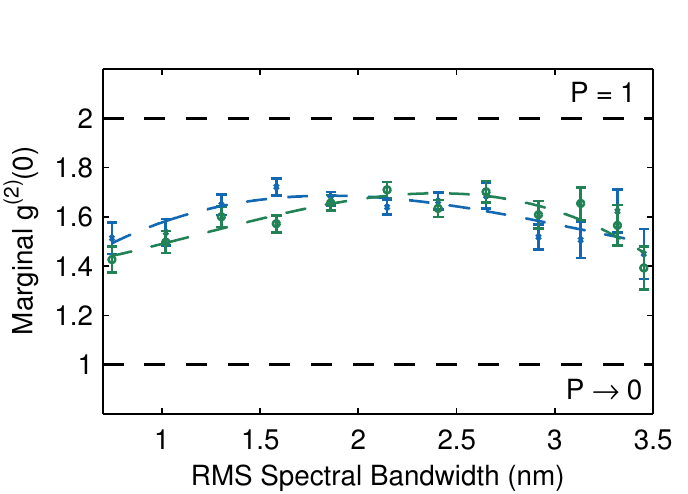}%{g2m/g2m_S1_and_S2_with_lines_curved_7x5_box}
        \caption{}
        \label{fig:g2m_S2_Results}
    \end{subfigure}
   \caption{(a)~Electron micrograph of cleaved PCF end face with hole-to-hole pitch of $1.49\,\mu$m and hole diameter of $d = 0.645\,\mu$m. (b)~Single source schematic. (c)~Simulated (left) and measured (right) two-photon joint spectral intensity for one source. (d)~Marginal second-order coherence measured for each source as a function of pump RMS spectral bandwidth.}
   \label{fig:Single_Source_Results}
\end{figure*}

The individual source design and performance is shown in Figure \ref{fig:Single_Source_Results}. We generated photon pairs by FWM in two lengths of photonic crystal fibre (PCF) dispersion-engineered to minimise frequency correlation between the photons by group-velocity matching, as detailed in the Supplementary Information~\cite{Garay-Palmett2007Photon-Pair-State-Preparation,Soller2010Bridging-Visible-and-Telecom}. The PCF was pumped with a 1064\,nm, 10\,MHz modelocked fibre laser (Fianium FemtoPower 1060-PP) to produce photon pairs at highly nondegenerate wavelengths: 810\,nm to enable efficient heralding with room-temperature silicon detectors, and 1550\,nm to access the low-loss telecoms window in commercial fibre. Fibre Bragg gratings (3\,dB bandwidth of 28\,nm around 1064\,nm) and a fibre wavelength division multiplexer were used to remove the pump pulses and separate the daughter photons. These provided at least 90\,dB of pump rejection in the signal and idler channels.

% fit to guide eye in g(2) plot

However, to isolate the signal and idler from noise outside the FBG rejection band (for example from Raman scattering and unwanted FWM processes) whilst also preserving the spectrum of the photon pairs we developed broadband filters based on photonic bandgap fibre (PBGF). We fabricated PBGFs with transmission bands centred at 800\,nm (PBGF-800) and 1550\,nm (PBGF-1550) to provide low in-band loss around the transmission wavelengths combined with high loss for out-of-band noise photons. Custom fibre tapers were incorporated to mode-match the PBGF-1550 with telecoms single-mode fibre (SMF) and minimise loss. Hence the outputs at 800\,nm and 1550\,nm were delivered to in conventional SMF (SM-800 and SMF-28 respectively) with total attenuation through all components following the PCF of $-5.6$~dB at 800~nm and $-7.0$~dB at 1550~nm (see Supplementary Information).

To confirm both the similarity of the two sources and the level of frequency correlation within each, the joint spectral intensity of the photon pairs, $|f(\omega_s, \omega_i)|^2$, was measured using stimulated emission tomography~\cite{Liscidini2013Stimulated_Emission_Tomography,Eckstein2014High_Resolution_Spectral} (SET; see Methods). The results for one source are shown in Fig.~\ref{fig:JSI_Upper}. The Schmidt decomposition of $|f(\omega_s, \omega_i)|$ placed an upper bound on the purity of the heralded single photons of $P < 0.86$, and the spectral overlap of the two sources was 95\%.

A direct measurement of the heralded single photon purity was obtained from the marginal second-order coherence, $g^{(2)}_{m}(0) = 1 + 1/K = 1 + P$. Two-mode squeezed states are characterised by thermal photon statistics in which each mode has $g^{(2)}_{m}(0) = 2$ leading to a heralded single-photon purity of $P = 1$~\cite{Eckstein2011Highly_Efficient_Single_Pass}; as the number of modes (and associated frequency correlation) increases the statistics become Poissonian and $g^{(2)}_{m}(0) \rightarrow 1$. The results illustrated in Fig.~\ref{fig:g2m_S2_Results} demonstrate that by matching the pump bandwidth with the phasematching function to minimise spectral correlation yielded a maximum value of $g^{(2)}_{m}(0) = 1.7 \pm 0.03$ corresponding to $P = 0.7 \pm 0.03$. We believe this to be limited primarily by the characteristics of our pump pulses.

%\subsection*{Source multiplexing}

\begin{figure*}
	\centering
    \begin{subfigure}[b]{12cm}
    		\centering
        \includegraphics[width = \textwidth]{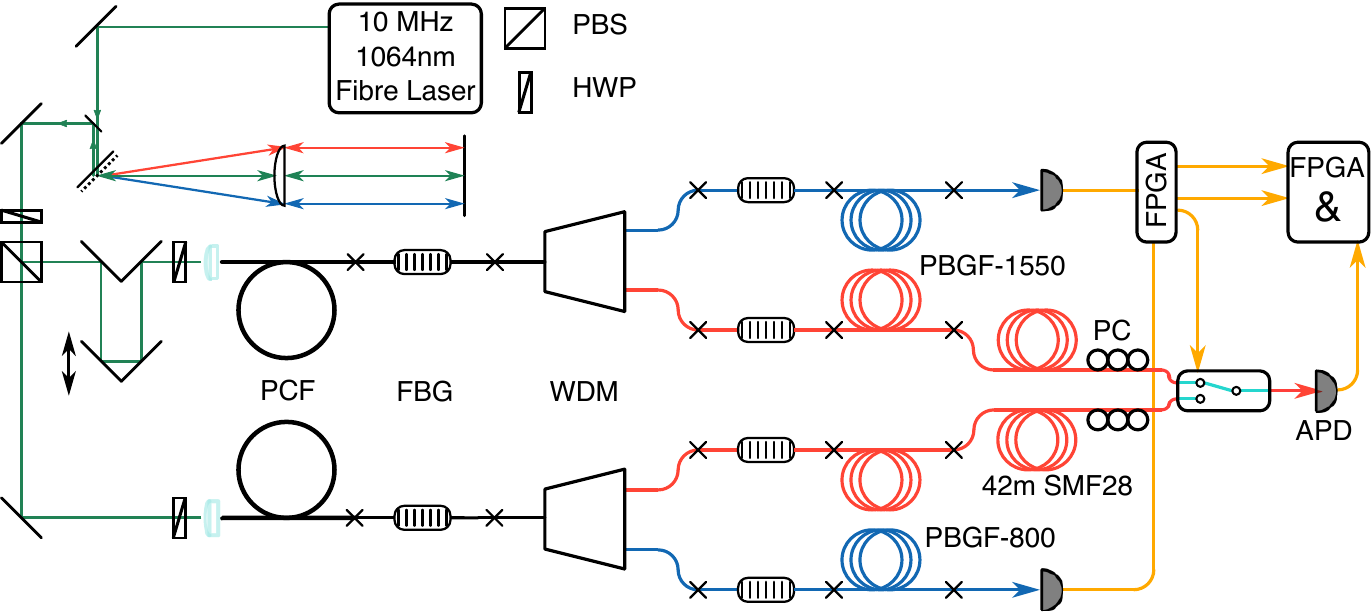}%{Multi_Source_Schematic_Coinc_Exp_Redesign4}
        \caption{}
		\label{fig:Multi_Source}
        \end{subfigure}
        
	\begin{subfigure}[b]{7cm}
    	\centering
        \includegraphics[width = \textwidth]{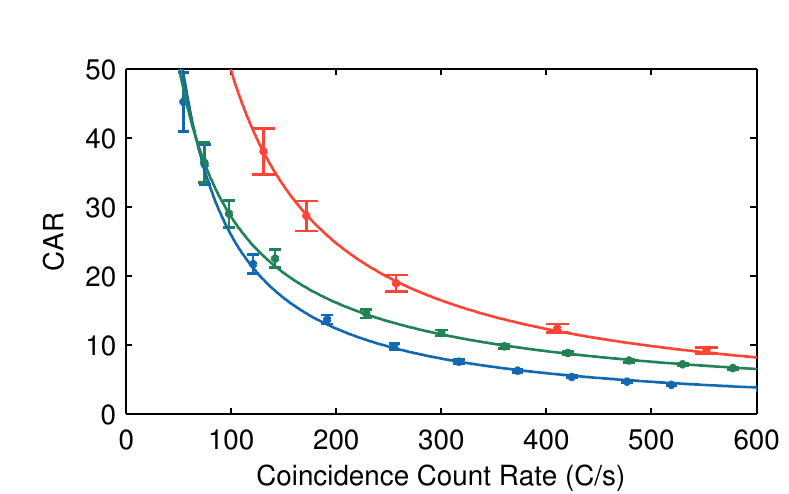}%{CAR/CARm_PowerLawFit_wo_Avg_8x5_box}
        \caption{}
        \label{fig:CAR_vs_CR}
   \end{subfigure}
   \begin{subfigure}[b]{7cm}
   		\centering
        \includegraphics[width = \textwidth]{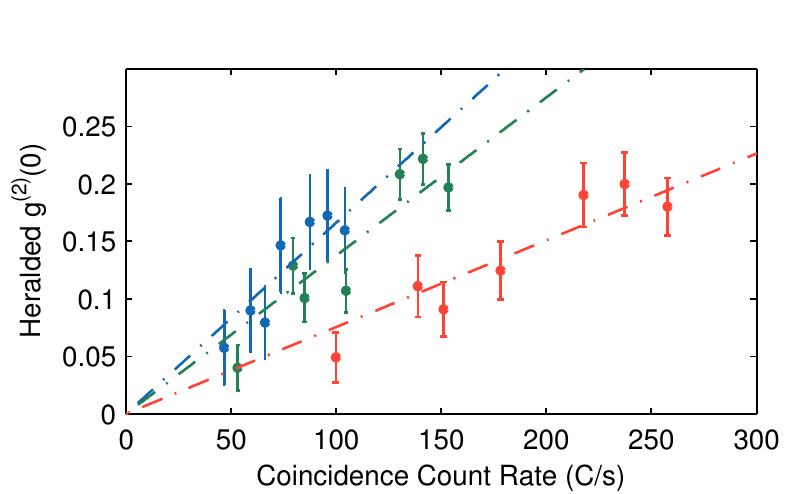}%{g2h/g2h_Narrow_Scan_wo_Avg_8x5_box2}
        \caption{}
        \label{fig:g2H_vs_CR}
  	\end{subfigure}
    \caption{(a)~Schematic of multiplexed source; note that the only alignment required is of the pump pulses into the two sections of PCF. (b)~Coincidence-to-accidental ratio and (c)~heralded second-order coherence as a function of coincidence count rate for individual sources (blue, green) and multiplexed source (red).}
  	\label{fig:Multiplexed_Results}
\end{figure*}

To multiplex the sources, the heralded output of each source was delayed by 200\,ns using 42\,m of SMF28 as shown in Figure \ref{fig:Multi_Source}. The delay lines were spliced to a $2 \times 1$ fibre-coupled optical switch (bandwidth 1\,MHz, insertion loss approximately $1$\,dB) to route the heralded photons to a single output via an in-line fibre polariser. One pump beam contained an optical delay to match arrival times between idler photons from each source. The switch state was set by a field programmable gate array (FPGA) conditioned on the heralding detection events, and the switch output was monitored by an InGaAs single photon detector.

The single and coincidence count rates were measured after the output of the switch for each source individually and with multiplexing activated. From the raw count rates we evaluated the coincidence-to-accidentals ratio (CAR; see methods) to quantify source performance with and without multiplexing, as shown in Fig.~\ref{fig:CAR_vs_CR}.  We measured the heralded second-order coherence, $g^{(2)}_{H}(0)$, by adding a 50:50 fibre coupler to the output of the switch for both individual sources and the multiplexed system. The results are shown in Fig.~\ref{fig:g2H_vs_CR} and demonstrate $g^{(2)}_{H}(0)$ far below the classical limit of 1, as required for a single-photon source. The value of $g^{(2)}_{H}(0)$ is non-zero due to multi-pair contributions and residual noise photons that pass the filters.

The benefit of multiplexing can be seen clearly from  the results in Fig.~\ref{fig:g2H_vs_CR}. At a CAR of 40, activating the switch improved the heralded single-photon rate by a factor of 1.93 relative to each individual source, close to the ideal value of 2. At this point the multiplexed coincidence count rate was 125\,s$^{-1}$ with a heralded $g^\text{(2)}(0) < 0.1$, whereas the individual sources returned $g^\text{(2)}(0) \approx 0.2$ to achieve the same count rates. Hence our results demonstrate that the multiplexed source achieves an increase in the probability of delivering a single photon without a corresponding increase in the probability of multi-photon noise. Furthermore, multiplexing remains advantageous when taking into account the effect of switch insertion loss on the individual sources.

Future developments that would enhance the performance of our source include extra dispersion control of the pump pulses to improve the heralded single photon purity, as well as more flexible shaping of the pump spectrum. Additionally we are developing new PCF designs that we anticipate will reduce noise from unwanted nonlinear processes hence enabling higher intrinsic pair generation rates without additional noise. A significant advantage of PCF-based FWM sources is that low average pump power is required per fibre, hence additional sources could straightforwardly be multiplexed from the same pump laser. Finally, a switchable delay network would allow several modes to be multiplexed in the time domain to combine the output from consecutive pump pulses.

%\section*{Conclusion}

With advanced microstructured fibre technology we have generated spectrally uncorrelated single photons in a fully fibre-integrated architecture. Heralding high-purity photons in telecommunication fibre modes gave us direct access to low-loss delay, enabling feed-forward to a fast optical switch. As a result, we have successfully multiplexed two fibre sources of high-purity photons and demonstrated an improvement in performance over similar individual sources. Our unique architecture requiring neither alignment nor narrow spectral filtering is amenable to multiplexing larger numbers of high-purity sources, providing a route to  high-performance single-photon sources capable of delivering large numbers of independent photons simultaneously.

\bibliographystyle{mosley_thesis}
\bibliography{BibDeskBibliography.bib}

\begin{thebibliography}{10}

\bibitem{Pan2012Multiphoton_Entanglement}
J.-W. Pan, Z.-B. Chen, C.-Y. Lu, H.~Weinfurter, A.~Zeilinger, and
  M.~\ifmmode~\dot{Z}\else \.{Z}\fi{}ukowski, \emph{Multiphoton entanglement
  and interferometry}, Rev. Mod. Phys., \textbf{84}, pp. 777--838 (2012).

\bibitem{Ladd2010Quantum_Computers}
T.~D. Ladd, F.~Jelezko, R.~Laflamme, Y.~Nakamura, C.~Monroe, and J.~L.
  O/'Brien, \emph{Quantum computers}, Nature, \textbf{464}, pp. 45--53 (2010).

\bibitem{Grice2001Eliminating-Frequency-and-Space-Time}
W.~P. Grice, A.~B. U'Ren, and I.~A. Walmsley, \emph{Eliminating frequency and
  space-time correlations in multiphoton states}, Phys. Rev. A, \textbf{64}, p.
  063815 (2001).

\bibitem{Burnham1970Observation-of-Simultaneity}
D.~C. Burnham and D.~L. Weinberg, \emph{Observation of Simultaneity in
  Parametric Production of Optical Photon Pairs}, Phys. Rev. Lett.,
  \textbf{25}, pp. 84--87 (1970).

\bibitem{Li2004All-Fiber-Photon-Pair-Sources}
X.~Li, J.~Chen, P.~Voss, J.~Sharping, and P.~Kumar, \emph{All-fiber photon-pair
  source for quantum communications: Improved generation of correlated
  photons}, Opt. Express, \textbf{12}, pp. 3737--3744 (2004).

\bibitem{Santori2002Indistinguishable_Photons_From}
C.~Santori, D.~Fattal, J.~Vuckovic, G.~S. Solomon, and Y.~Yamamoto,
  \emph{Indistinguishable photons from a single-photon device}, Nature,
  \textbf{419}, pp. 594--597 (2002).

\bibitem{Englund2010Deterministic_Coupling_of_a}
D.~Englund, B.~Shields, K.~Rivoire, F.~Hatami, J.~Vu{\v c}kovi{\'c}, H.~Park,
  and M.~D. Lukin, \emph{Deterministic Coupling of a Single Nitrogen Vacancy
  Center to a Photonic Crystal Cavity}, Nano Letters, \textbf{10}, pp.
  3922--3926 (2010), pMID: 20825160.

\bibitem{Nilsson2013Quantum_Teleportation_Using}
NilssonJ., S.~M., C.~H. A., Skiba-SzymanskaJ., LucamariniM., W.~B., B.~J.,
  S.~L., FarrerI., R.~A., and S.~J., \emph{Quantum teleportation using a
  light-emitting diode}, Nat Photon, \textbf{7}, pp. 311--315 (2013).

\bibitem{Christ2012Limits-on-the-Deterministic}
A.~Christ and C.~Silberhorn, \emph{Limits on the deterministic creation of pure
  single-photon states using parametric down-conversion}, Phys. Rev. A,
  \textbf{85}, p. 023829 (2012).

\bibitem{Yao2012Observation-of-Eight-Photon}
X.-C. Yao, T.-X. Wang, P.~Xu, H.~Lu, G.-S. Pan, X.-H. Bao, C.-Z. Peng, C.-Y.
  Lu, Y.-A. Chen, and J.-W. Pan, \emph{Observation of eight-photon
  entanglement}, Nat Photon, \textbf{6}, pp. 225--228 (2012).

\bibitem{Nunn2013Enhancing-Multiphoton-Rates}
J.~Nunn, N.~K. Langford, W.~S. Kolthammer, T.~F.~M. Champion, M.~R. Sprague,
  P.~S. Michelberger, X.-M. Jin, D.~G. England, and I.~A. Walmsley,
  \emph{Enhancing Multiphoton Rates with Quantum Memories}, Phys. Rev. Lett.,
  \textbf{110}, p. 133601 (2013).

\bibitem{Migdall2002Tailoring-Single-Photon}
A.~L. Migdall, D.~Branning, and S.~Castelletto, \emph{Tailoring single-photon
  and multiphoton probabilities of a single-photon on-demand source}, Phys.
  Rev. A, \textbf{66}, p. 053805 (2002).

\bibitem{Adam2014Optimisation_of_Periodic}
P.~Adam, M.~Mechler, I.~Santa, and M.~Koniorczyk, \emph{Optimization of
  periodic single-photon sources}, Phys. Rev. A, \textbf{90}, p. 053834 (2014).

\bibitem{Bonneau2014Effect_of_loss_on}
D.~{Bonneau}, G.~J. {Mendoza}, J.~L. {O'Brien}, and M.~G. {Thompson},
  \emph{{Effect of Loss on Multiplexed Single-Photon Sources}}, ArXiv e-prints
  (2014).

\bibitem{Francis-Jones2014Exploting_The_Limits}
R.~J.~A. {Francis-Jones} and P.~J. {Mosley}, \emph{Exploring the limits of
  multiplexed photon-pair sources for the preparation of pure single-photon
  states}, ArXiv e-prints (2014).

\bibitem{Ma2011Experimental_Generation_of}
X.-s. Ma, S.~Zotter, J.~Kofler, T.~Jennewein, and A.~Zeilinger,
  \emph{Experimental generation of single photons via active multiplexing},
  Phys. Rev. A, \textbf{83}, p. 043814 (2011).

\bibitem{Collins2013Integrated_Spatial_Multiplexing}
M.~J. Collins, C.~Xiong, I.~H. Rey, T.~D. Vo, J.~He, S.~Shahnia, C.~Reardon,
  T.~F. Krauss, M.~J. Steel, A.~S. Clark, and B.~J. Eggleton, \emph{Integrated
  spatial multiplexing of heralded single-photon sources}, Nat Commun,
  \textbf{4} (2013).

\bibitem{Meany2014Hybrid-photonic-circuit}
T.~Meany, L.~A. Ngah, M.~J. Collins, A.~S. Clark, R.~J. Williams, B.~J.
  Eggleton, M.~J. Steel, M.~J. Withford, O.~Alibart, and S.~Tanzilli,
  \emph{Hybrid photonic circuit for multiplexed heralded single photons}, Laser
  and Photonics Reviews, \textbf{8}, pp. L42--L46 (2014).

\bibitem{Kaneda2015Time_Multiplexed_Heralded}
F.~Kaneda, B.~G. Christensen, J.~J. Wong, H.~S. Park, K.~T. McCusker, and P.~G.
  Kwiat, \emph{Time-multiplexed heralded single-photon source}, Optica,
  \textbf{2}, pp. 1010--1013 (2015).

\bibitem{Mendoza2016Active_Tempora_and_Spatial}
G.~J. Mendoza, R.~Santagati, J.~Munns, E.~Hemsley, M.~Piekarek,
  E.~Mart\'{i}n-L\'{o}pez, G.~D. Marshall, D.~Bonneau, M.~G. Thompson, and
  J.~L. O'Brien, \emph{Active temporal and spatial multiplexing of photons},
  Optica, \textbf{3}, pp. 127--132 (2016).

\bibitem{Silverstone2014On-Chip-Quantum-Interference}
S.~W., BonneauD., OhiraK., SuzukiN., YoshidaH., IizukaN., EzakiM., N.~M.,
  T.~G., H.~H., ZwillerV., M.~D., R.~G., O.~L., and T.~G., \emph{On-chip
  quantum interference between silicon photon-pair sources}, Nat Photon,
  \textbf{8}, pp. 104--108 (2014).

\bibitem{Mosley2008Heralded-Generation-of-Ultrafast}
P.~J. Mosley, J.~S. Lundeen, B.~J. Smith, P.~Wasylczyk, A.~B. U'Ren,
  C.~Silberhorn, and I.~A. Walmsley, \emph{Heralded Generation of Ultrafast
  Single Photons in Pure Quantum States}, Phys. Rev. Lett., \textbf{100}, p.
  133601 (2008).

\bibitem{Halder2009Nonclassical_Two_Photon}
M.~Halder, J.~Fulconis, B.~Cemlyn, A.~Clark, C.~Xiong, W.~J. Wadsworth, and
  J.~G. Rarity, \emph{Nonclassical 2-photon interference with separate
  intrinsically narrowband fibre sources}, Opt. Express, \textbf{17}, pp.
  4670--4676 (2009).

\bibitem{Cohen2009Tailored-Photon-Pair-Generation}
O.~Cohen, J.~S. Lundeen, B.~J. Smith, G.~Puentes, P.~J. Mosley, and I.~A.
  Walmsley, \emph{Tailored Photon-Pair Generation in Optical Fibers}, Phys.
  Rev. Lett., \textbf{102}, p. 123603 (2009).

\bibitem{Soller2011High-Performance-Single-Photon}
C.~S\"oller, O.~Cohen, B.~J. Smith, I.~A. Walmsley, and C.~Silberhorn,
  \emph{High-performance single-photon generation with commercial-grade optical
  fiber}, Phys. Rev. A, \textbf{83}, p. 031806 (2011).

\bibitem{Eckstein2011Highly_Efficient_Single_Pass}
A.~Eckstein, A.~Christ, P.~J. Mosley, and C.~Silberhorn, \emph{Highly Efficient
  Single-Pass Source of Pulsed Single-Mode Twin Beams of Light}, Phys. Rev.
  Lett., \textbf{106}, p. 013603 (2011).

\bibitem{Garay-Palmett2007Photon-Pair-State-Preparation}
K.~Garay-Palmett, H.~J. McGuinness, O.~Cohen, J.~S. Lundeen, R.~Rangel-Rojo,
  A.~B. U'ren, M.~G. Raymer, C.~J. McKinstrie, S.~Radic, and I.~A. Walmsley,
  \emph{Photon pair-state preparation with tailored spectral properties by
  spontaneous four-wave mixing in photonic-crystal fiber}, Opt. Express,
  \textbf{15}, pp. 14870--14886 (2007).

\bibitem{Soller2010Bridging-Visible-and-Telecom}
C.~S\"oller, B.~Brecht, P.~J. Mosley, L.~Y. Zang, A.~Podlipensky, N.~Y. Joly,
  P.~S.~J. Russell, and C.~Silberhorn, \emph{Bridging visible and telecom
  wavelengths with a single-mode broadband photon pair source}, Phys. Rev. A,
  \textbf{81}, p. 031801 (2010).

\bibitem{Liscidini2013Stimulated_Emission_Tomography}
M.~Liscidini and J.~E. Sipe, \emph{Stimulated Emission Tomography}, Phys. Rev.
  Lett., \textbf{111}, p. 193602 (2013).

\bibitem{Eckstein2014High_Resolution_Spectral}
A.~Eckstein, G.~Boucher, A.~Lema{\^\i}tre, P.~Filloux, I.~Favero, G.~Leo, J.~E.
  Sipe, M.~Liscidini, and S.~Ducci, \emph{High-resolution spectral
  characterization of two photon states via classical measurements}, Laser and
  Photonics Reviews, \textbf{8}, pp. L76--L80 (2014).

\bibitem{Love1986Quantifying_Loss_Minimisation}
J.~Love and W.~Henry, \emph{Quantifying loss minimisation in single-mode fibre
  tapers}, Electronics Letters, \textbf{22}, pp. 912--914 (1986).

\bibitem{Birks1992The_Shape_of_Fibre_Tapers}
T.~Birks and Y.~Li, \emph{The shape of fiber tapers}, Lightwave Technology,
  Journal of, \textbf{10}, pp. 432--438 (1992).

\bibitem{Jizan2015Bi_Photon_Spectral}
I.~Jizan, L.~G. Helt, C.~Xiong, M.~J. Collins, D.-Y. Choi, C.~Joon~Chae,
  M.~Liscidini, M.~J. Steel, B.~J. Eggleton, and A.~S. Clark, \emph{Bi-photon
  spectral correlation measurements from a silicon nanowire in the quantum and
  classical regimes}, Scientific Reports, \textbf{5}, pp. 12557 EP -- (2015).

\bibitem{Luan2004All_Solid_Photonic_Bandgap_Fibre}
F.~Luan, A.~K. George, T.~D. Hedley, G.~J. Pearce, D.~M. Bird, J.~C. Knight,
  and P.~S.~J. Russell, \emph{All-solid photonic bandgap fiber}, Opt. Lett.,
  \textbf{29}, pp. 2369--2371 (2004).

\bibitem{Pearce2005Adaptive_Curvilinear_Coordinates}
G.~J. Pearce, T.~D. Hedley, and D.~M. Bird, \emph{Adaptive curvilinear
  coordinates in a plane-wave solution of Maxwell's equations in photonic
  crystals}, Phys. Rev. B, \textbf{71}, p. 195108 (2005).

\bibitem{Stone2006An_Improved_Photonic}
J.~M. Stone, G.~J. Pearce, F.~Luan, T.~A. Birks, J.~C. Knight, A.~K. George,
  and D.~M. Bird, \emph{An improved photonic bandgap fiber based on an array of
  rings}, Opt. Express, \textbf{14}, pp. 6291--6296 (2006).

\end{thebibliography}
%\bibliography{160318_multiplexing_arXiv.bbl}
%\bibliography{160318_multiplexing_pjm.bbl}

\subsection*{Acknowledgements}

We gratefully acknowledge the fibre tapering expertise of S Yerolatsitis and T Birks and technical assistance from S Renshaw. We acknowledge financial support from the UK EPSRC through the First Grant programme (EP/K022407/1) and through the UK Quantum Technology Hub \textit{Networked Quantum Information Technologies} (EP/M013243/1).

\subsection*{Author contributions}

RJAFJ and PJM designed the experiment. RJAFJ and RAH fabricated the fibre and developed the electronics. RJAFJ carried out the experiment and analysed the data. PJM supervised the project. All authors participated in discussions and contributed to the manuscript.

\subsection*{Additional information}

Supplementary information accompanies this submission. Correspondence and requests for materials should be addressed to RJAFJ.

\subsection*{Competing financial interests}

The authors declare no competing financial interests.

\subsection*{Methods}

\textbf{Source Fabrication}. The PCF and PBGF were fabricated at the University of Bath using the stack-and-draw technique. The PCF dispersion was fine tuned by controlling the hole-to-hole pitch via the fibre diameter and the hole size varied by pressurising the holes in the fibre preform. To create fibre tapers for mode matching, a series of large-mode conventional fibres were fabricated and then post-processed on a flame-brush taper-rig. These were designed to transform adiabatically the mode of SMF28 (mode-field diameter (MFD) $\sim 9\,\mu$m) to that of the PBGF-1550 (MFD $\sim 18\,\mu$m)\cite{Love1986Quantifying_Loss_Minimisation,Birks1992The_Shape_of_Fibre_Tapers}. The sources were assembled by fusion splicing the components in a conventional arc splicer.
\\
\\
\textbf{Stimulated Emission Tomography}. The JSI distribution of the photon-pairs generated in each PCF source was measured using stimulated four-wave mixing \cite{Liscidini2013Stimulated_Emission_Tomography,Jizan2015Bi_Photon_Spectral}. The PCF was pumped at 1064~nm and the FWM stimulated by a CW seed laser (bandwidth $<$125\,kHz, tuning range 1520 -- 1630\,nm) that was tuned across the range of idler wavelengths. The resulting stimulated FWM signal was recorded with an optical spectrum analyser to find the distribution of signal and idler emission wavelengths proportional to $|f(\omega_s, \omega_i)|^2$.
\\
\\
\textbf{Coincidence-to-Accidentals Ratio}.
The coincidence-to-accidental ratio for a single source, $\text{CAR}^{(1)}$, was calculated by taking the ratio of the measured coincidence count rate, $N_{c}^{(1)}$, to the accidental coincidence rate, $A_{c}^{(1)}$:
\begin{equation}
	\text{CAR}^{(1)} = \frac{N_{c}^{(1)}}{A_{c}^{(1)}} = \frac{R_p N_{c}^{(1)}}{N_{H}^{(1)} N_{I}^{(1)}},
\end{equation}
where $N_{H}^{(1)}$ and $N_{I}^{(1)}$ are the single count rates in the herald and idler arms respectively and $R_p$ is the repetition rate of the laser.

To calculate the coincidence-to-accidental ratio for our multiplexed source, $\text{CAR}^{(2)}$, we first found the total number of distinct coincidence counts per second, $N_{c}^{(2)}$, and the total accidentals count rate $A_{c}^{(2)}$,
\begin{eqnarray}
	N_{c}^{(2)} & = & N_{H_{1}I} + N_{H_{2}I} - N_{H_{1}H_{2}I},\\
    A_{c}^{(2)} & = & \frac{N_{H_{1}}N_{I}}{R_{p}} + \frac{N_{H_{2}}N_{I}}{R_{p}} - \frac{N_{H_{1}}N_{H_{2}}N_{I}}{R_{p}^{2}},
\end{eqnarray}
The subtraction of the third term on the RHS prevents the double counting of events when both sources produce a heralding signal, but the heralded photon from source 1 is blocked at the switch. The two source $\text{CAR}^{(2)}$ is then,
\begin{equation}
	\text{CAR}^{(2)} = \frac{N_{c}^{(2)}}{A_{c}^{(2)}}.
\end{equation}
\\
\\
\textbf{Second Order Coherence Functions.}
To measure the second-order coherence functions of the idler photons from each source, a 50:50 fibre coupler was incorporated after the switch and in-line polariser, and an additional InGaAs detector monitored the second output of the fibre coupler. The two detectors were gated by the pump laser train and synchronised. The coincidence count rate was measured with an integration time of between 15 and 45 mins. From the raw coincidence count rates we evaluated the marginal second-order coherence function,
\begin{equation}
	g^{(2)}_{m}(0) = \frac{N_{I_{1}I_{2}}\cdot R_{p}}{N_{I_{1}}\cdot N_{I_{2}}},
\end{equation}
or heralded second-order coherence
\begin{equation}
	g^{(2)}(0) = \frac{N_{H I_{1}I_{2}}\cdot R_{p}}{N_{H I_{1}}\cdot N_{H I_{2}}}.
\end{equation}
Note that for all figures in this manuscript no background subtraction has been carried out.

\section*{Supplementary Information - All-fibre multiplexed source of high-purity heralded single photons}
\setcounter{figure}{0}
\renewcommand{\figurename}{Supplementary Figure}
\renewcommand{\thefigure}{S\arabic{figure}}
\subsection*{PCF Dispersion Engineering}

Elimination of spectral correlation between signal and idler photons results in a joint spectral amplitude (JSA), $f(\omega_{s},\omega_{i})$, that can be factorised: $f(\omega_{s},\omega_{i}) = g_{s}(\omega_{s})\cdot g_{i}(\omega_{i})$. Each function $g_{j}(\omega_{j})$ is solely dependent on either the signal or idler photon.

Two effects contribute to correlation in the JSA. Energy conservation results in anti-correlation through the pump envelope function $\alpha(\omega_{s} + \omega_{i})$, and momentum conservation is described by the phasematching function $\phi(\omega_{s},\omega_{i})$ which is dependent on the dispersion of the medium. The JSA is given by the product of these two functions :$f(\omega_{s},\omega_{i}) = \alpha(\omega_{s} + \omega_{i}) \phi(\omega_{s},\omega_{i})$.  Hence by orientating the phasematching function such that the correlation it imparts on the JSA completely counters the contribution of the pump envelope, we can produce a two-photon state in which there are no spectral correlations. In practice this can be achieved by setting the group velocity of the pump to be between that of the signal and the idler \cite{Grice2001Eliminating-Frequency-and-Space-Time, Garay-Palmett2007Photon-Pair-State-Preparation}.

\begin{figure}[!b]
	\centering
    \begin{subfigure}[b]{0.49\textwidth}
    \centering
		\includegraphics[width = \textwidth]{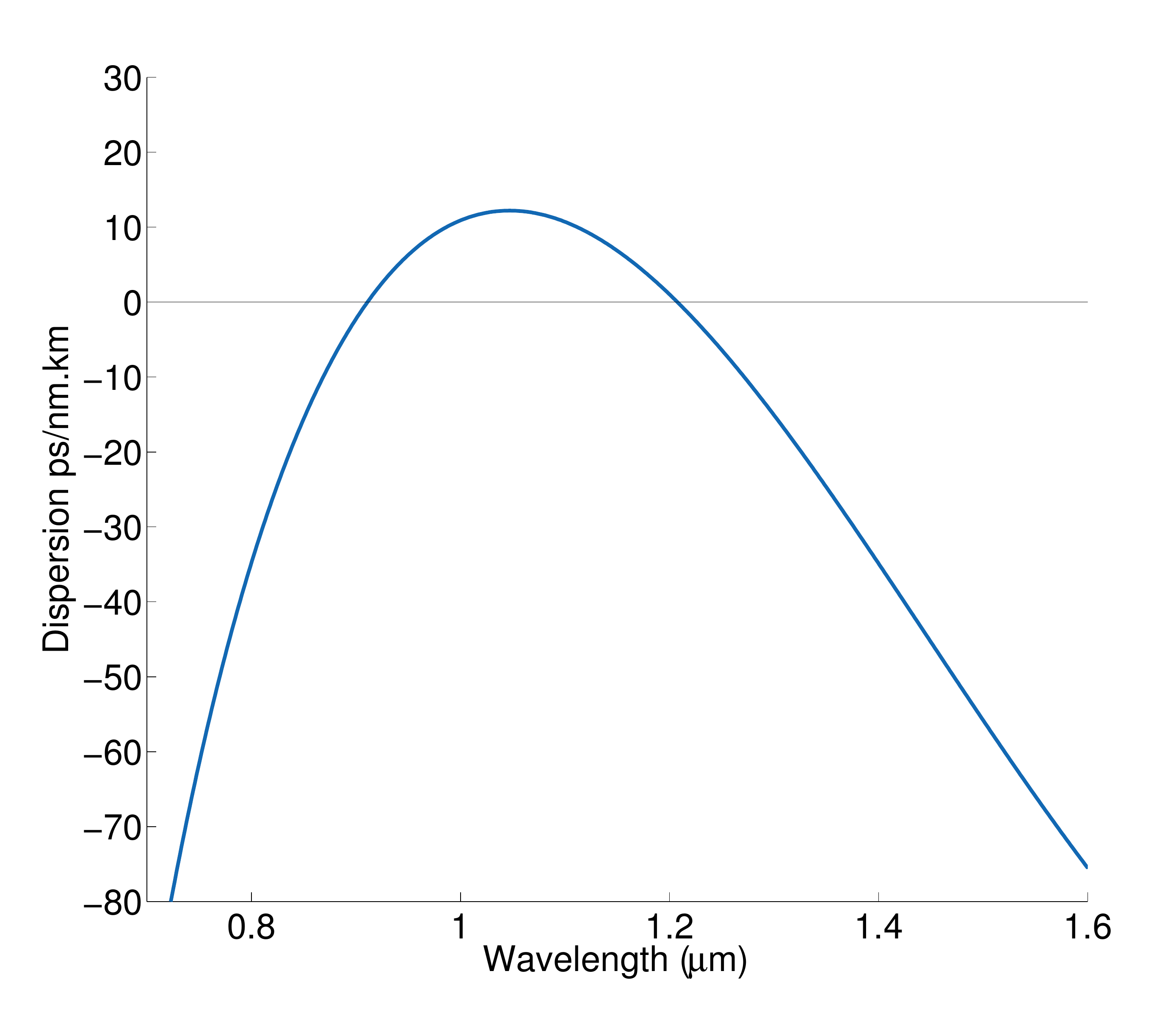}
    	\caption{}
        \label{fig:Sim_GVD}
    \end{subfigure}
    \begin{subfigure}[b]{0.49\textwidth}
    	\centering
        \includegraphics[width = \textwidth]{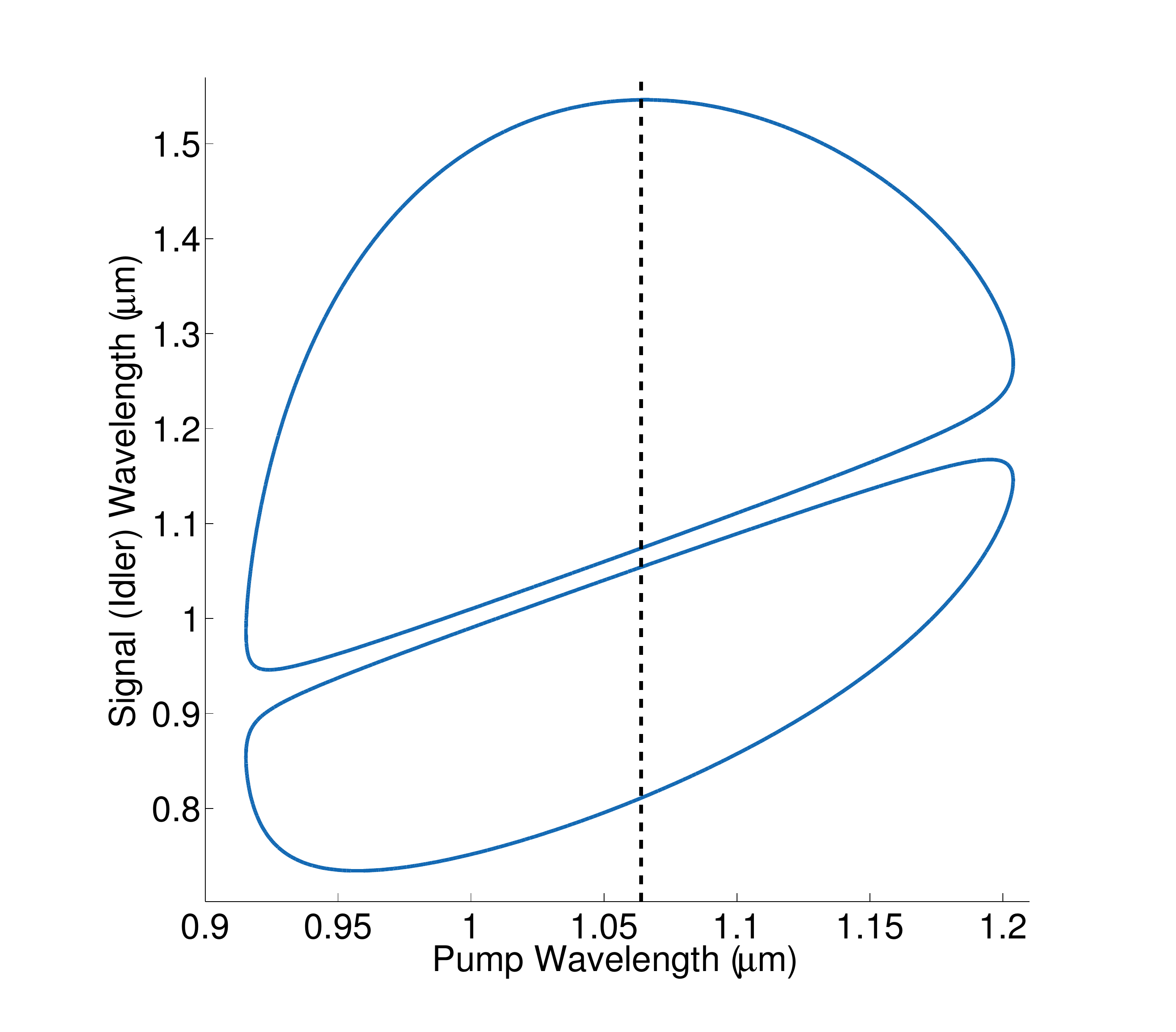}
        \caption{}
        \label{fig:sim_PM_Contours}
	\end{subfigure}
	
	\begin{subfigure}[b]{0.49\textwidth}
    \centering
		\includegraphics[width = \textwidth]{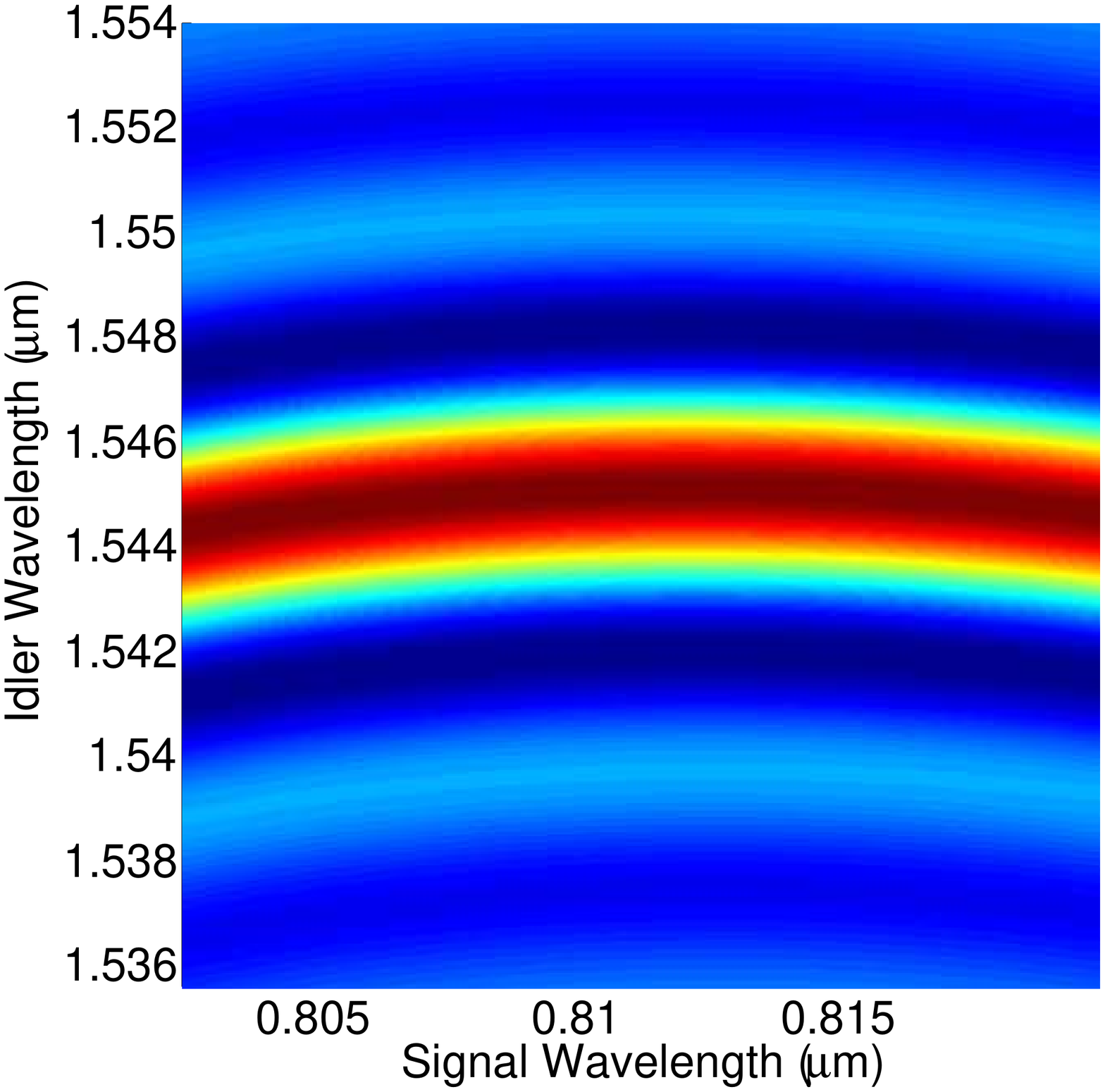}
    	\caption{}
        \label{fig:Sim_PMF}
    \end{subfigure}
    \begin{subfigure}[b]{0.49\textwidth}
    	\centering
        \includegraphics[width = \textwidth]{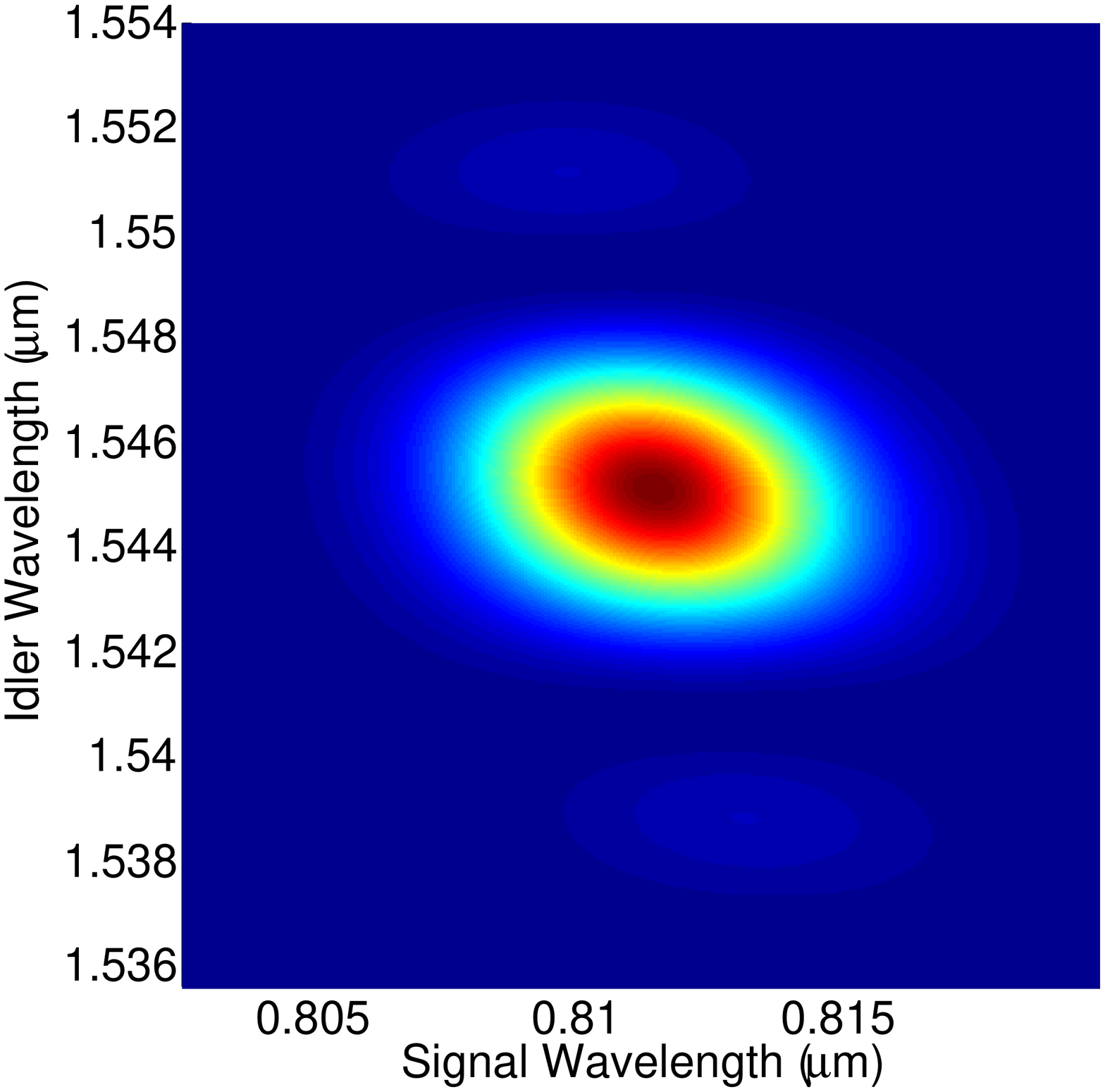}
        \caption{}
        \label{fig:Sim_JSI}
	\end{subfigure}
	\caption{(a)~Simulated group-velocity dispersion and (b)~FWM phasematching contours. (c)~Simulated phasematching function. (d)~Simulated joint spectral intensity.}
\end{figure}

\begin{figure}
    	\centering
    	\includegraphics[width = 0.8\textwidth]{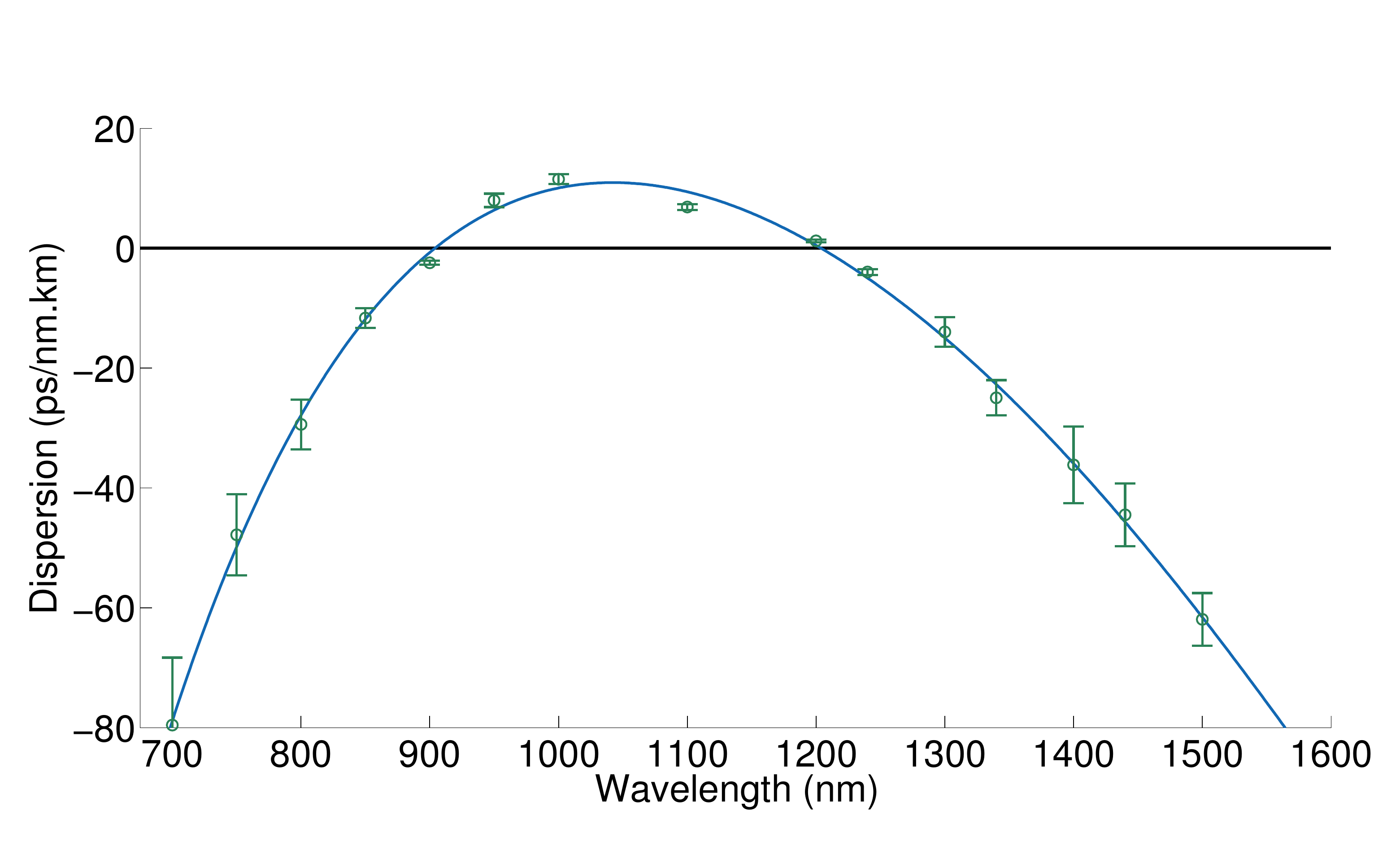}
        \caption{~Group-velocity dispersion measured by white-light interferometry.}
        \label{fig:Meas_GVD}
\end{figure}

PCF gives the ability to tune group-velocity dispersion through control of the fibre cladding. We designed and fabricated a PCF with two zero dispersion wavelengths as seen in Fig.~\ref{fig:Sim_GVD} and \ref{fig:Meas_GVD}. The resulting phasematching contours, the loci of points with zero phase mismatch between the pump, signal and idler are shown in Fig.~\ref{fig:sim_PM_Contours}. The two ZDWs give phasematching contours that form a pair of closed loops, one for the signal and one for the idler. Hence by tuning the pump wavelength, the correlation induced in the two-photon JSA by the phasematching function may be controlled, and used to negate the correlation imparted by the pump envelope function.

The closed loop phasematching contour means that factorability will always be satisfied for a certain range of pump wavelengths, but the phasematched signal and idler wavelengths may not be ideal. Nevertheless, through careful design the PCF structure can be tuned to achieve minimal correlation for a set of target FWM wavelengths, in this case 800\,nm and 1550\,nm. We have targeted a state where the signal becomes group-velocity matched to the pump. In this case the signal photon is localised to the pump temporal mode, whereas the idler photon walks off from the pump, yielding a narrower bandwidth idler photon. This corresponds to the phasematching function lying parallel to the signal axis in Fig.~\ref{fig:Sim_PMF}, when the bandwidths of the phasematching function and pump envelope function are matched, the JSI depicted in Fig.~\ref{fig:Sim_JSI} is formed.

\begin{figure}
	\centering
    \includegraphics[width = 0.8\textwidth]{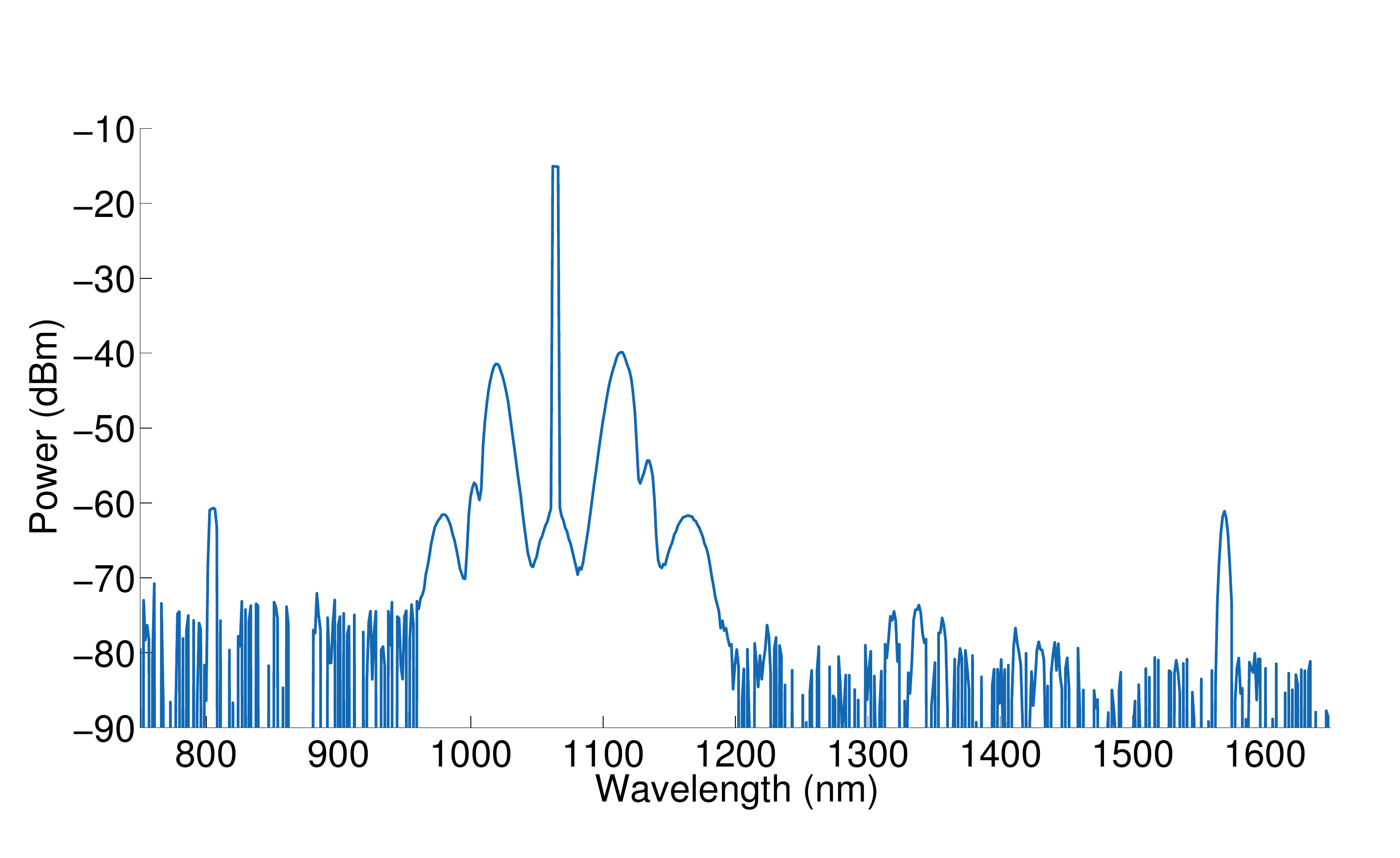}
    \caption{Bright FWM spectrum of PCF pumped with microchip laser.}
    \label{fig:Bright_Light_FWM}
\end{figure}
During the fibre drawing process, the fabricated fibre was characterised with bright FWM using a 1064nm nanosecond pulsed microchip laser to confirm the wavelength of phasematched FWM. An example FWM spectrum is shown in Fig.~\ref{fig:Bright_Light_FWM}. The fibre draw parameters were then adjusted to tune the FWM peaks to the target wavelengths. In Fig.~\ref{fig:Bright_Light_FWM} the outer-sideband phasematched solutions at 800\,nm and 1550\,nm can be clearly seen. Closer to the pump, cascaded FWM can be seen as a result of inner-sideband phasematching; these inner-sideband processes are one source of noise that our PBGF filters had to be capable of removing.

% Include phasematching function, Pump Function JSI plots from thesis

\subsection*{Photonic Bandgap Fibre Filtering}

\begin{figure}
	\centering
	  \begin{subfigure}[b]{0.6\textwidth}
    \includegraphics[width = \textwidth]{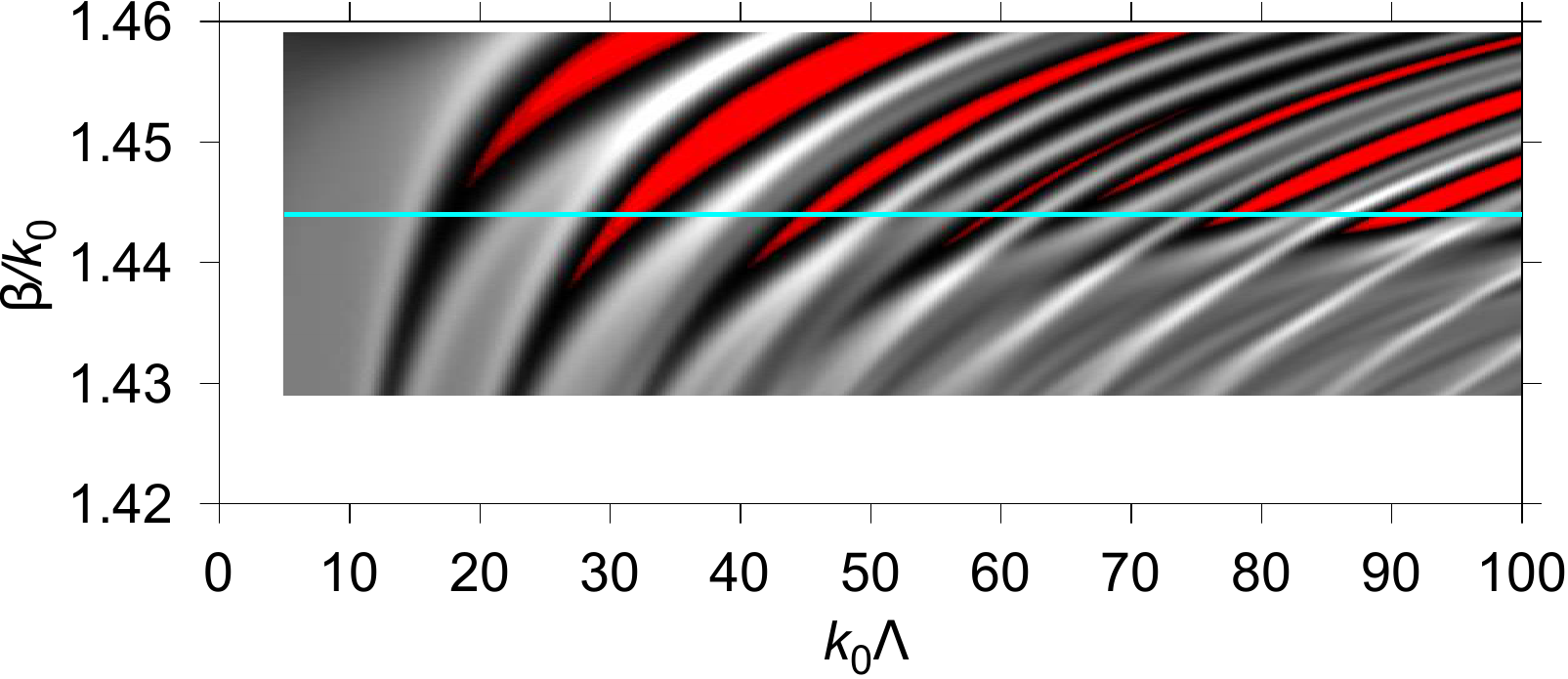}
    \caption{}
    \end{subfigure}
	\centering
    \begin{subfigure}[b]{0.49\textwidth}
    	\centering
        \includegraphics[width = \textwidth]{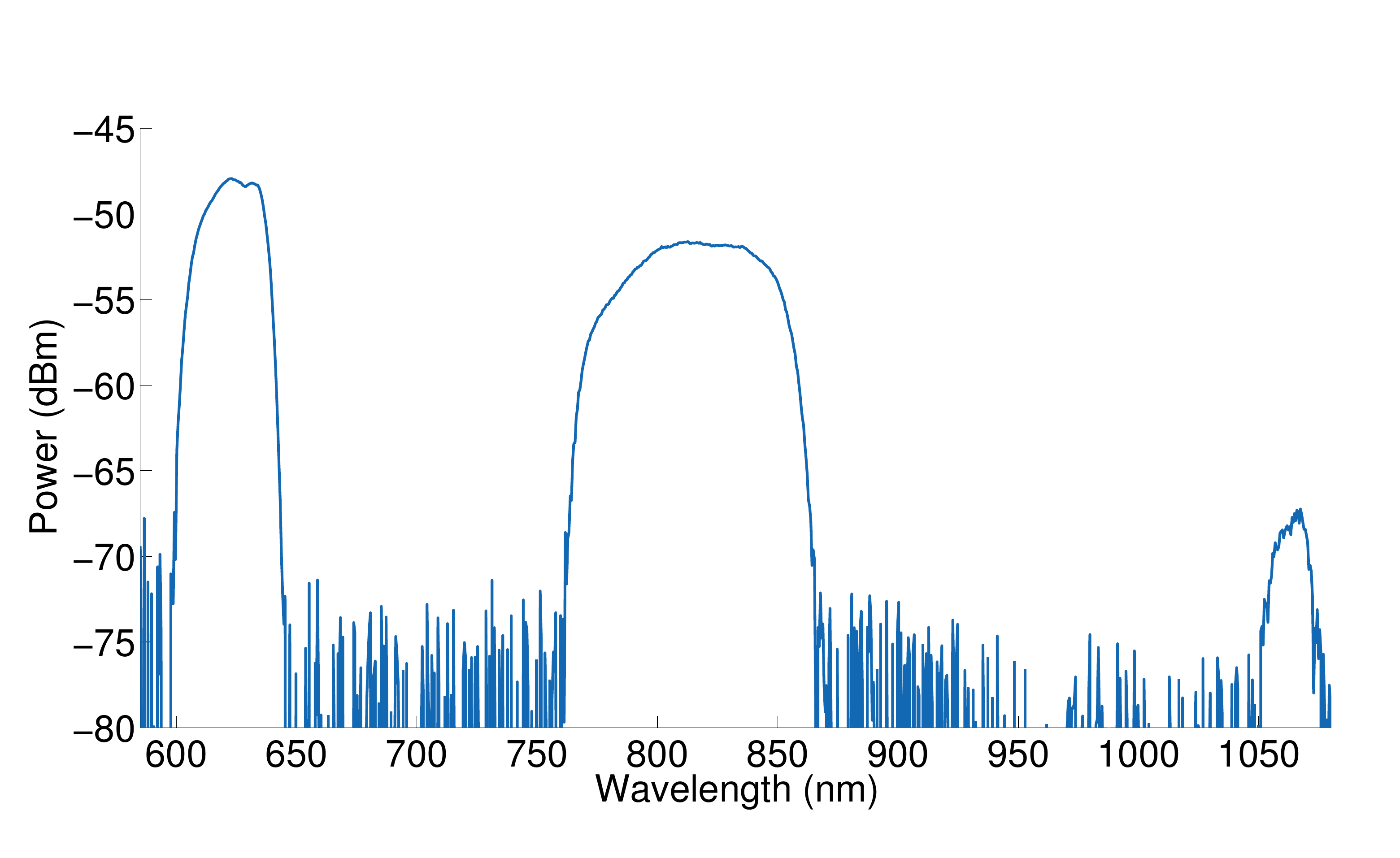}
        \caption{}
	\end{subfigure}
    \begin{subfigure}[b]{0.49\textwidth}
    	\centering
        \includegraphics[width = \textwidth]{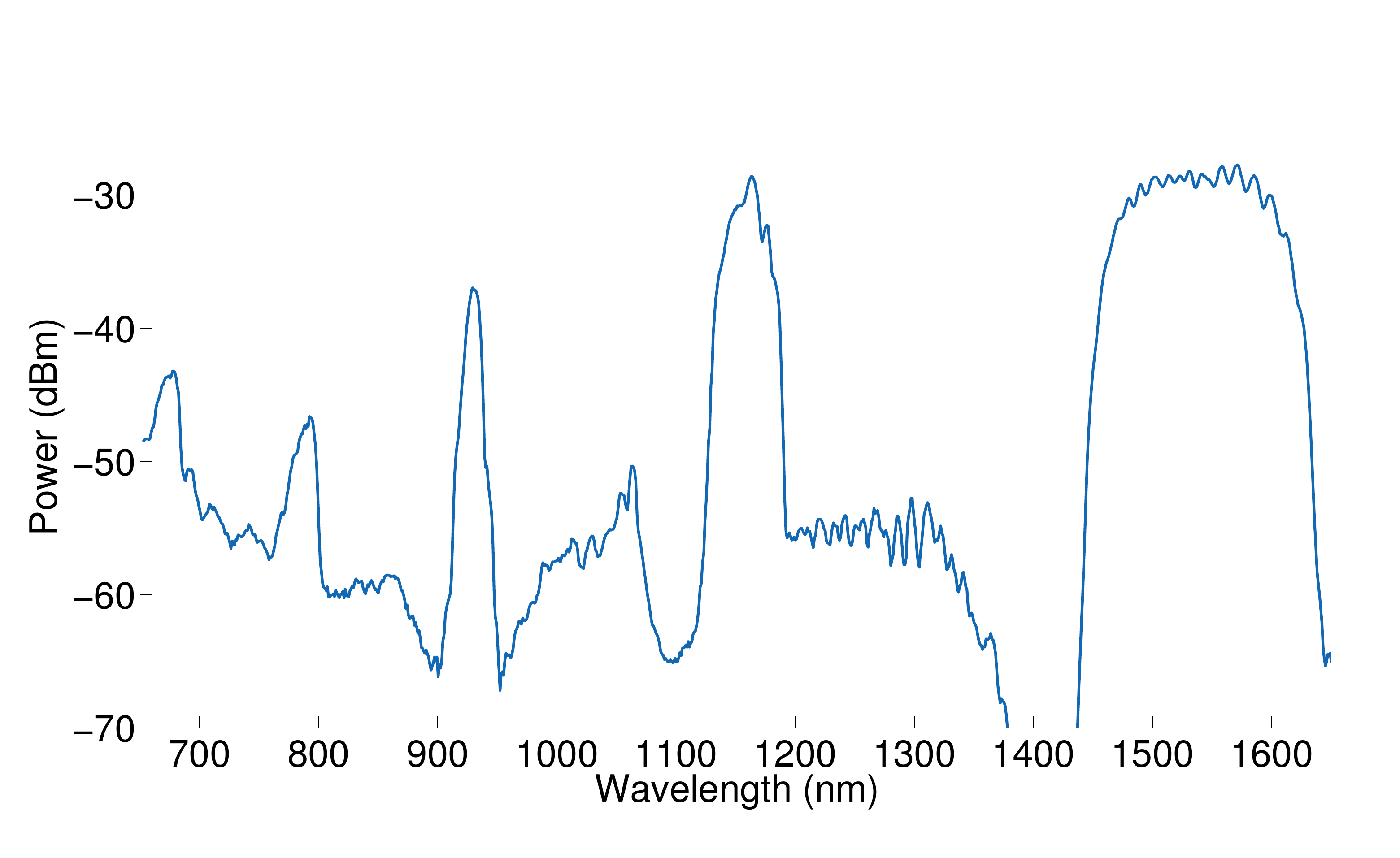}
		\caption{}
	\end{subfigure}
	\caption{(a) Photonic density of states simulation for PBGF cladding. (b) PBGF-800 transmission spectrum. (c) PBGF-1550 transmission spectrum.}
    \label{fig:PBGF_Transmission}
\end{figure}

Figure~\ref{fig:PBGF_Transmission} shows the photonic density of states for photonic bandgap fibre (PBGF) cladding consisting of a triangular lattice of high refractive index inclusions (Ge-doped rods) in a background of fused silica, calculated using the fixed-frequency plane wave method~\cite{Luan2004All_Solid_Photonic_Bandgap_Fibre, Pearce2005Adaptive_Curvilinear_Coordinates}. The effective refractive index is plotted as a function of free-space wavevector normalised to the pitch of the cladding. Grey regions correspond to a high density of states where the cladding supports sets of guided modes, while regions shown in red correspond to the photonic bandgaps of the cladding in which guided modes are not supported in the cladding. The blue line corresponds to the refractive index of the target wavelength, 1550\,nm, in silica, and where it intersects the red regions, bandgaps are formed in the cladding at the target wavelength. Any light with a wavevector lying within the bandgap cannot propagate in the cladding, and therefore becomes trapped within the low index core~\cite{Stone2006An_Improved_Photonic}. Figure~\ref{fig:PBGF_Transmission} displays the resulting transmission spectra of PBGF-800 and PBGF-1550, measured using a supercontinuum light source.

\begin{figure}
	\centering
	\begin{subfigure}[b]{0.8\textwidth}
    	\centering
        \includegraphics[width = 0.8\textwidth]{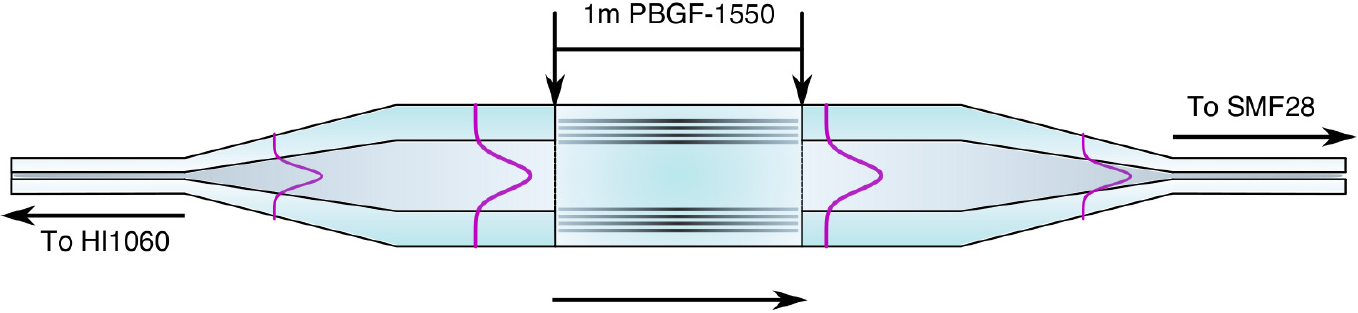}
        \caption{}
        \label{fig:PBGF_Filter}
	\end{subfigure}
    
    \begin{subfigure}[b]{0.8\textwidth}
		\centering
    	\includegraphics[width = 0.8\textwidth]{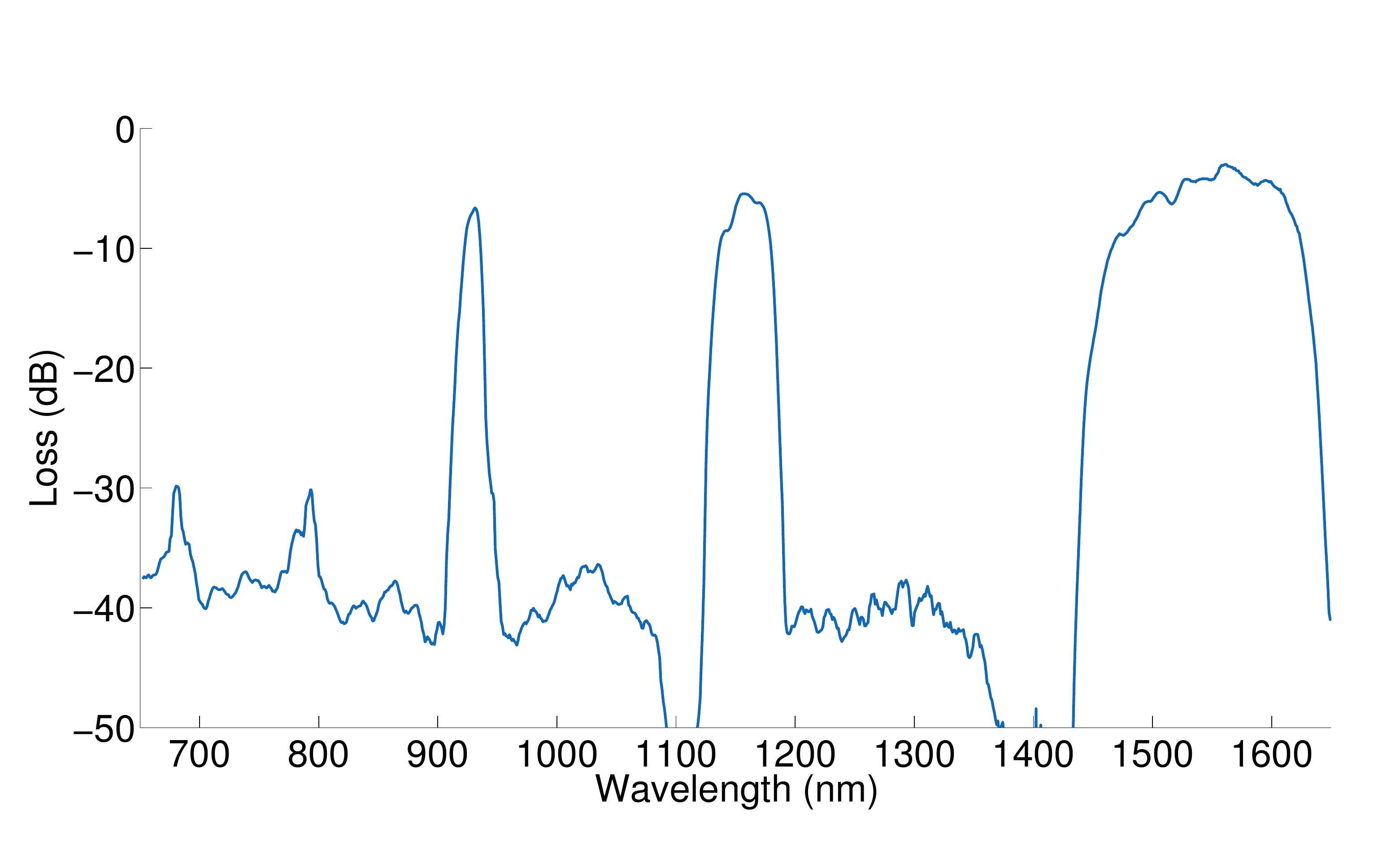}
    	\caption{}
        \label{fig:PBGF_Filter_Loss}
	\end{subfigure}
    \caption{(a) Schematic of filter with taper transitions. (b) Transmission loss through completed PBGF-1550 Filter.}
\end{figure}

Due to the large pitch required to form bandgaps at the correct wavelength for PBGF-1550, the guided mode was significantly larger than that of SMF-28 resulting in poor splice performance. To reduce the loss and improve the mechanical stability of the splice between PBGF-1550 and SMF-28, a series of tapered-fibre transitions were fabricated using the flame-brush technique~\cite{Birks1992The_Shape_of_Fibre_Tapers}. A graded-index large mode area (LMA) fibre was fabricated in which the guided mode and the outer diameter were both matched to that of PBGF-1550 to ensure low loss and high mechanical strength when spliced. The LMA fibre was tapered to a diameter of 125um to match that of SMF-28 whilst also providing good mode field overlap. A schematic of the complete spliced PBGF-1550 filter is shown in Fig.~\ref{fig:PBGF_Filter}. The fundamental mode entering from the SMF-28 was adiabatically transformed in the up-taper region and then launched into the PBGF-1550, and a second down-taper on the output side transformed the mode back to match that of SMF-28 \cite{Love1986Quantifying_Loss_Minimisation}. The transmission through the complete filter (3\,dB at 1550\,nm) was measured by performing a cut-back measurement and is shown in Fig.~\ref{fig:PBGF_Filter_Loss}.

\subsection*{Source transmission measurement}

\begin{figure}
    \begin{subfigure}[b]{\textwidth}
    	\centering
    	\includegraphics[width = 0.8\textwidth]{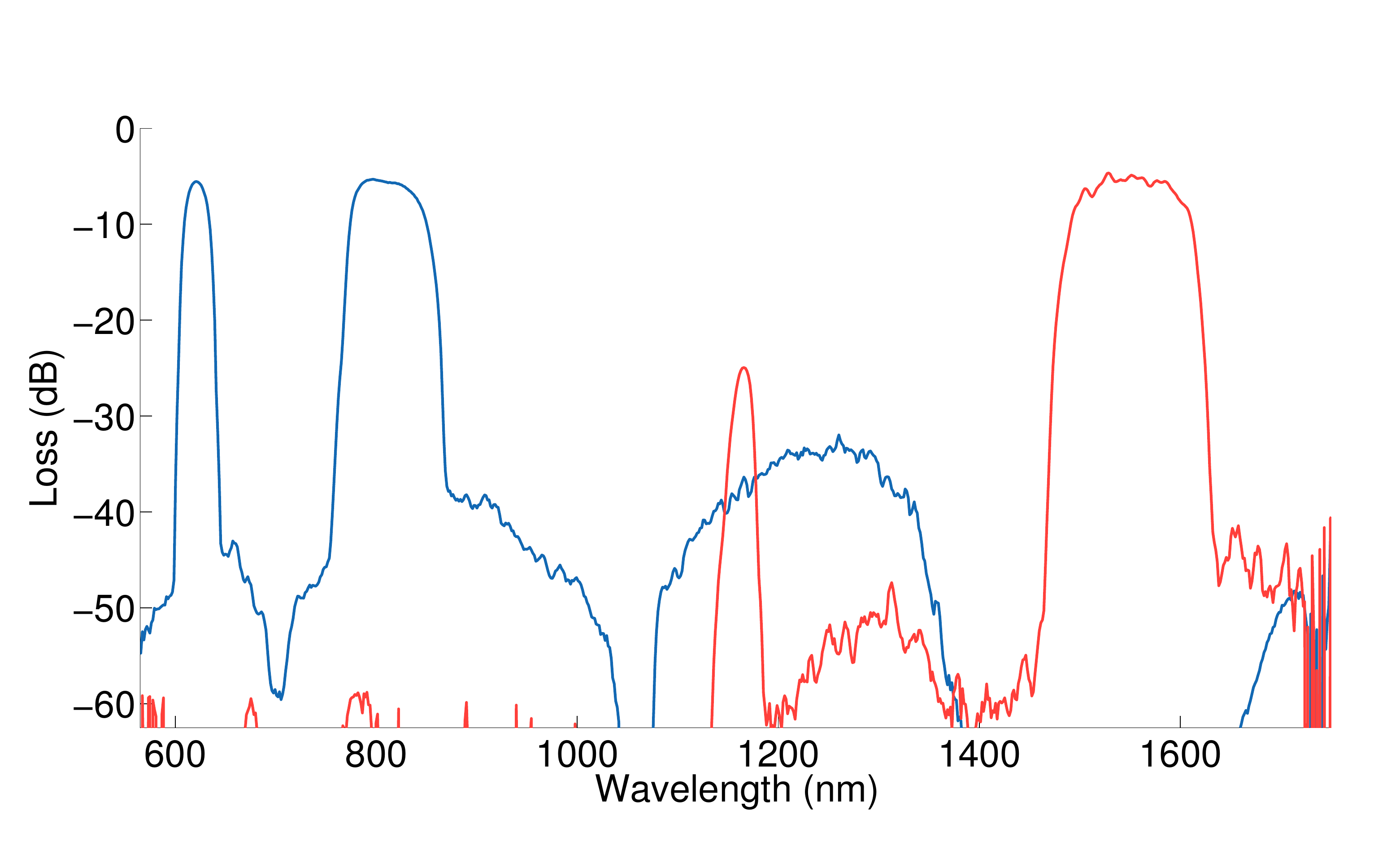}
	\end{subfigure}
	\caption{Measured transmission for completed source. Blue and red correspond to the signal and idler channels respectively.}
	 \label{fig:Transmission_Loss}
\end{figure}

A cut-back measurement was made using a supercontinuum light source to find the transmission of one complete photon-pair source, from the PCF to the output of the final single mode fibres. The spectrally-resolved transmission is shown in Fig.~\ref{fig:Transmission_Loss} for the signal and idler channels. The low-loss regions in each arm are largely defined by the bandgaps of the PBGF filters. The total loss was found to be 5.6dB and 5.0dB for the signal and idler respectively. In the idler arm following the single mode fibre, an extra 1dB was incurred due to the transmission loss of the optical switch used for multiplexing, and a further 1dB of loss in the in-line polariser.

\subsection*{Stimulated emission tomography of the joint spectral density}

The FWM joint spectral density was measured by stimulating the process with a seed laser at one of the daughter wavelengths, following the method proposed by Sipe and Liscidini known as stimulated emission tomography \cite{Liscidini2013Stimulated_Emission_Tomography}. As the spontaneous and stimulated forms of the nonlinear optical interaction are governed by the same phasematching and energy conservation, the stimulated signal is proportional to the probability amplitude of the two photon state.

\begin{figure}
	\centering
    \begin{subfigure}[b]{0.6\textwidth}
    	\centering
        \includegraphics[width = 0.9\textwidth]{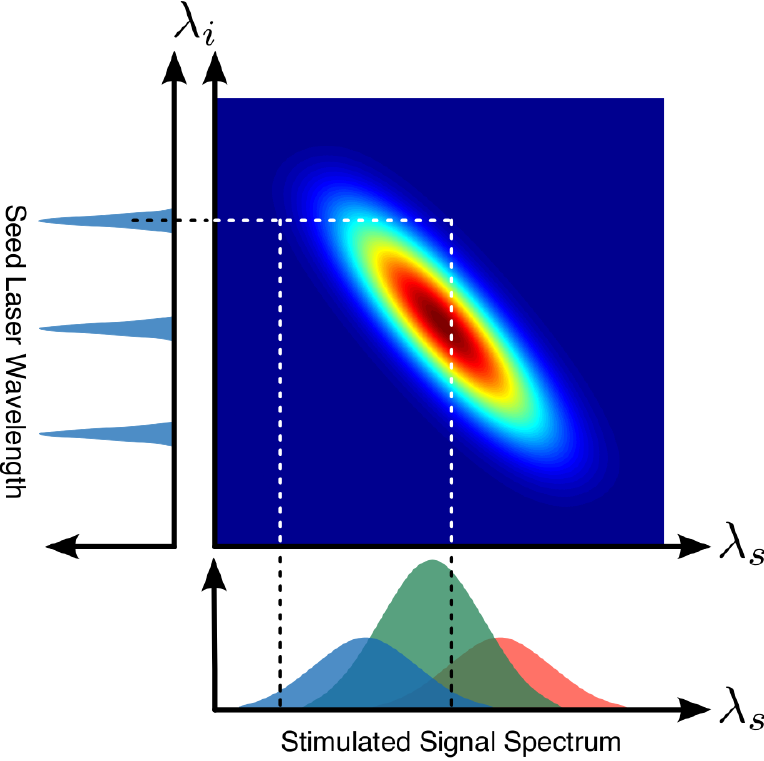}
        \caption{}
        \label{fig:Stim_JSI}
	\end{subfigure}
    \begin{subfigure}[b]{0.6\textwidth}
    	\centering
        \includegraphics[width = 0.9\textwidth]{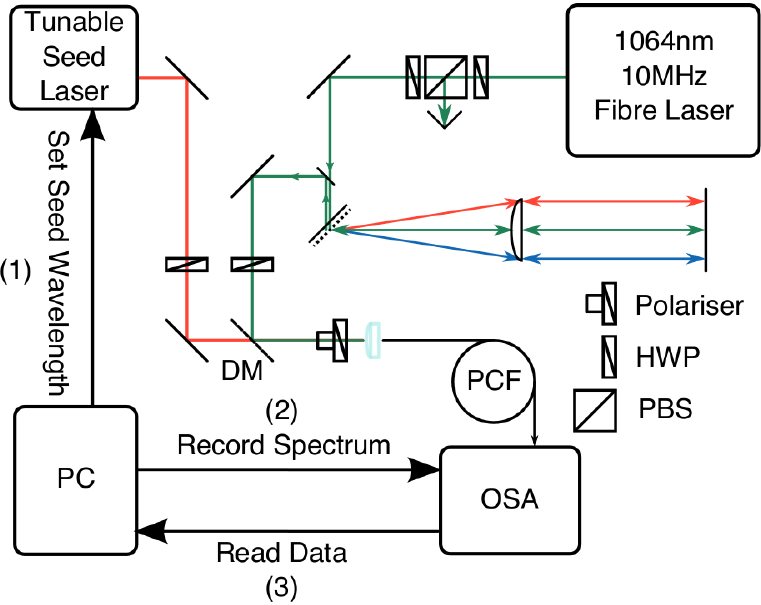}
        \caption{}
        \label{fig:Stim_Set_Up}
	\end{subfigure}
	
	\centering
    \begin{subfigure}[b]{0.32\textwidth}
    	\centering
        \includegraphics[width = \textwidth]{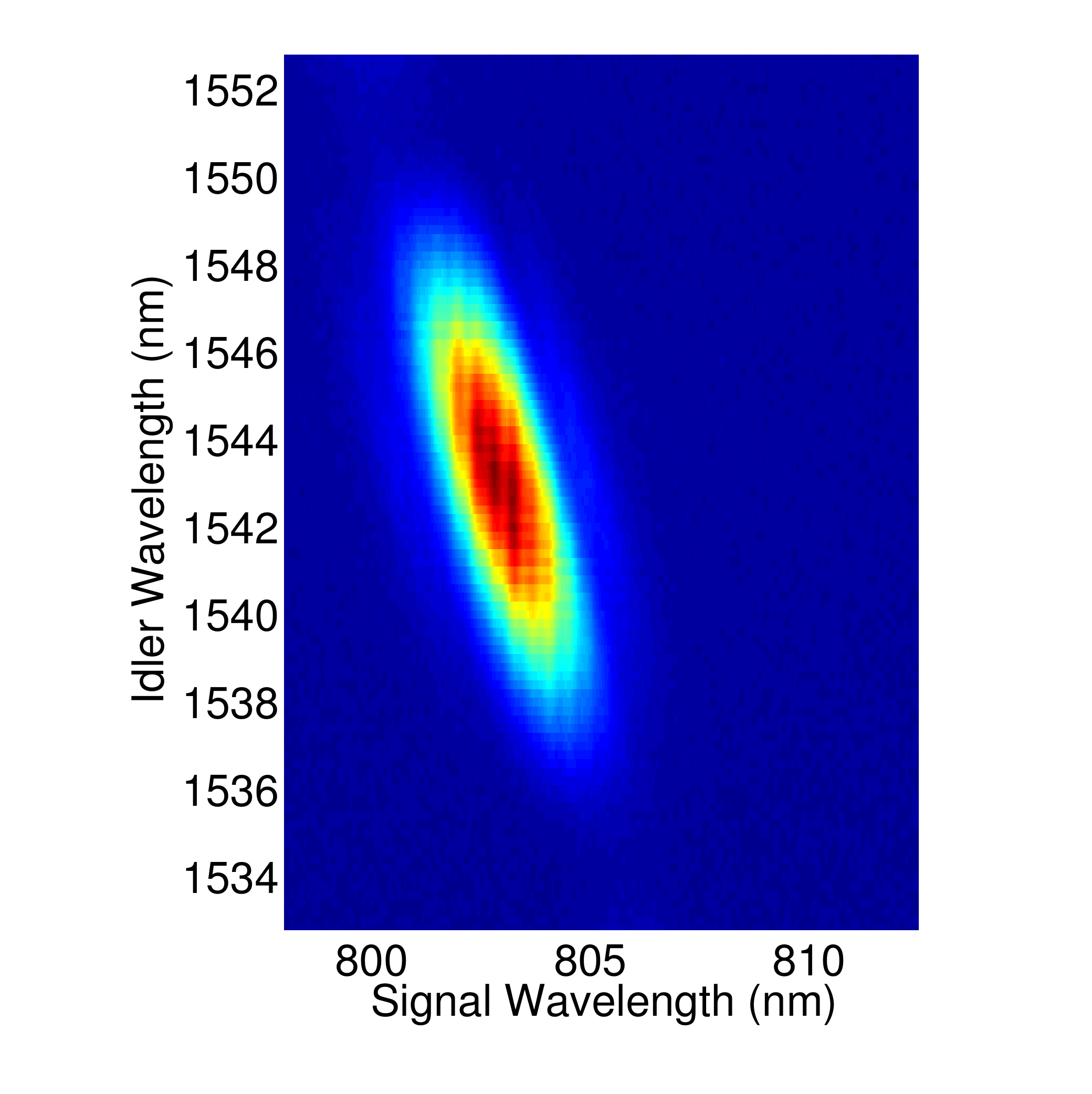}
        \caption{}
	\end{subfigure}
    \begin{subfigure}[b]{0.32\textwidth}
    	\centering
		\includegraphics[width = \textwidth]{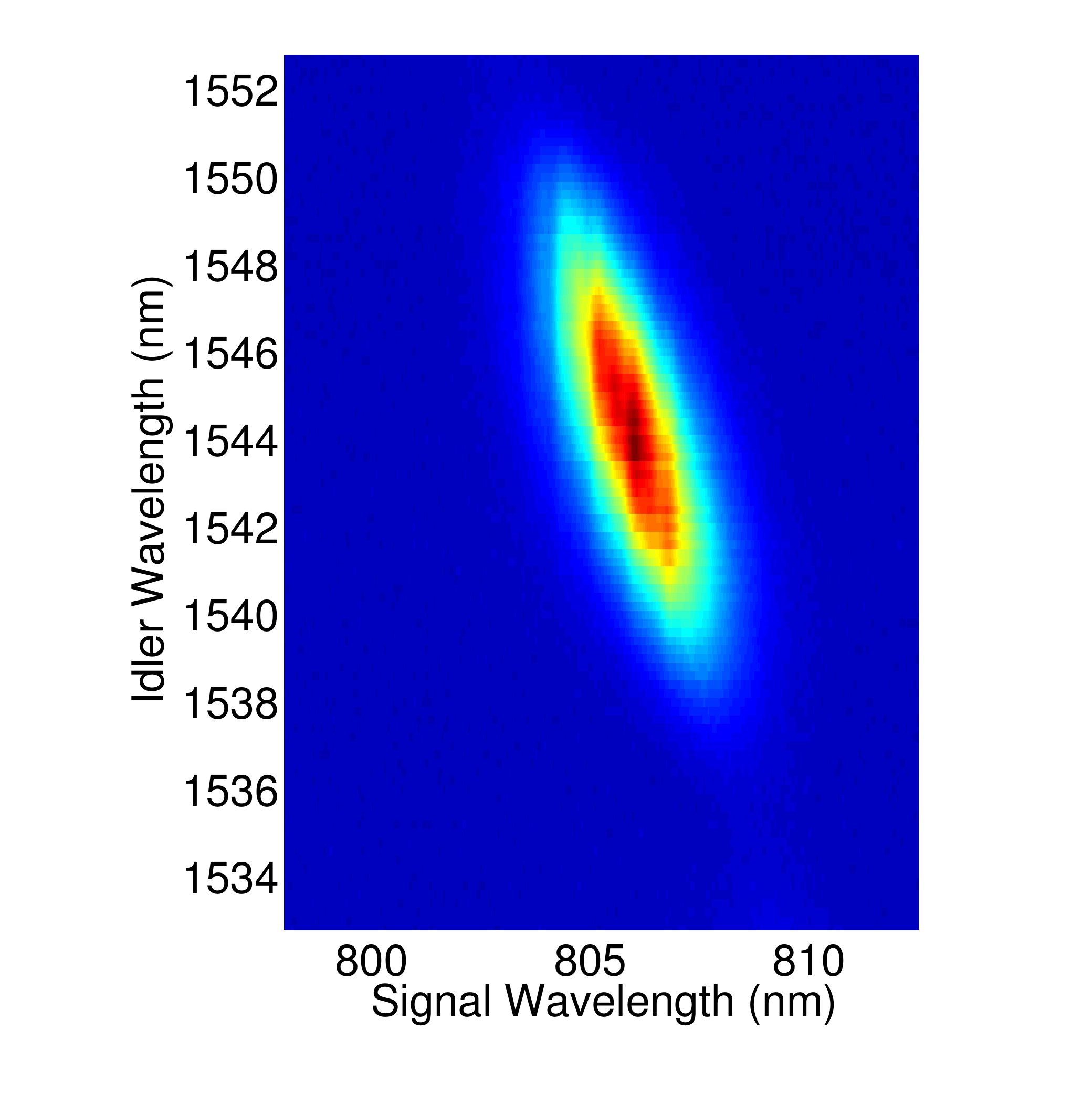}
        \caption{}
	\end{subfigure}
    \begin{subfigure}[b]{0.32\textwidth}
    	\centering
        \includegraphics[width = \textwidth]{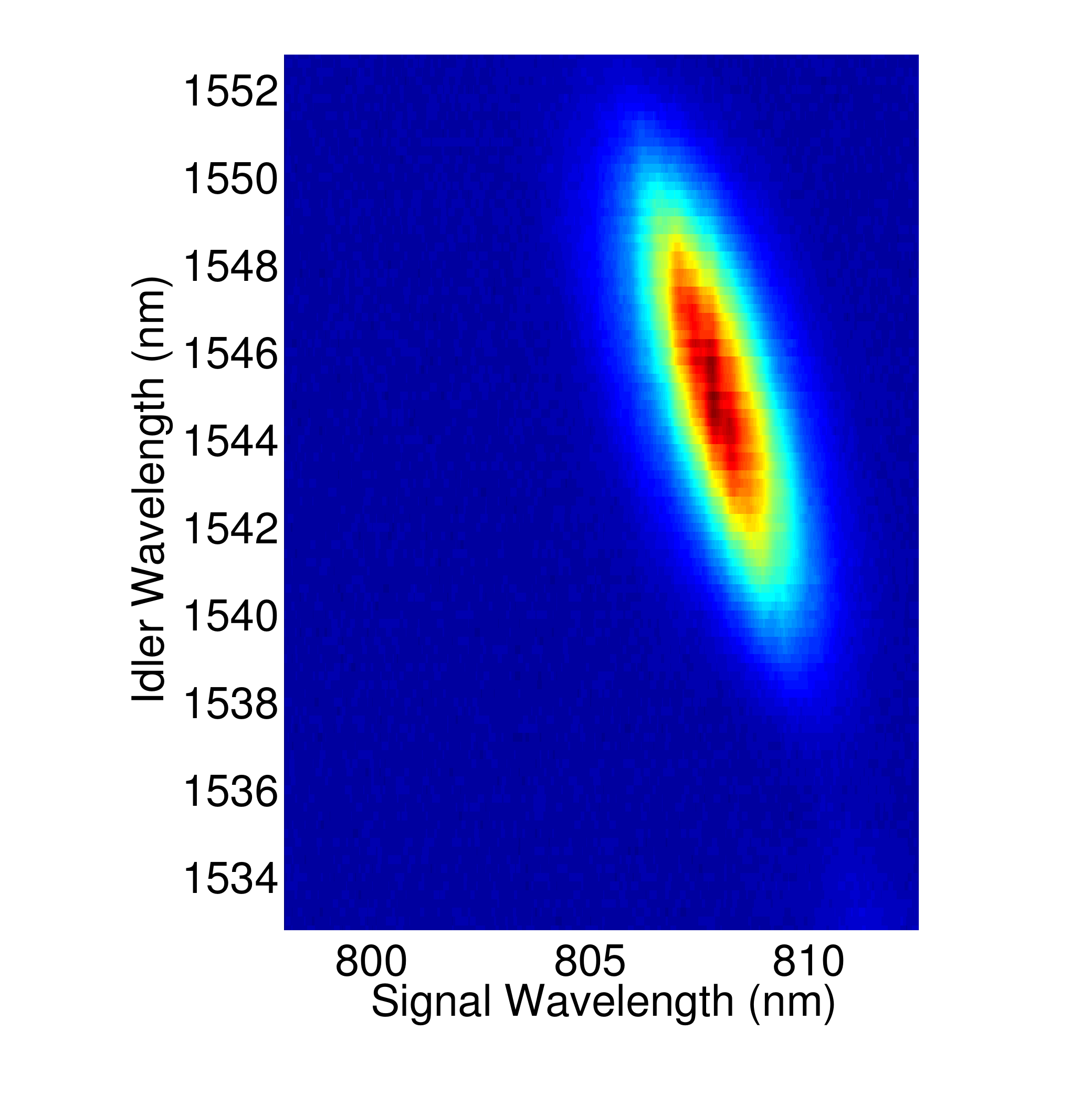}
        \caption{}
	\end{subfigure}
    \caption{(a)~Stimulated emission tomography measurement. (b)~Schematic of setup for stimulated emission tomography. (c--e)~Joint spectrum measured for three different pump wavelengths.}
    \label{fig:JSI_Wav_Cent}
\end{figure}

A schematic of the set-up used to measure the joint spectral intensity distribution is shown in Fig.~\ref{fig:Stim_Set_Up}. Each setting of the seed wavelength gave one stimulated signal spectrum as seen Fig.~\ref{fig:Stim_JSI}. By stacking these spectra the joint spectrum was recovered. The measurement can be made in a short time and at high resolution using an optical spectrum analyser rather than photon-counting detectors.

Figure~\ref{fig:JSI_Wav_Cent} illustrates the results for a single length of PCF at three different pump wavelengths. The bandwidths of the signal and idler photons are not equal as in this instance the bandwidth of the phasematching function (set by the fibre length) is not well-matched to that of the pump function. Together these three plots may be used to infer the orientation of the phasematching function. As the pump wavelength is tuned, the phasematched signal wavelength increases without a commensurate increase in the idler wavelength. This indicates that the phasematching function lies parallel to the signal axis. This is in agreement with our earlier simulation of the phasematching contours where the resulting signal photon was group-velocity matched to the pump. 

We have used this technique to match the bandwidth of the pump pulses to the fibre length, and hence produce photon pairs with minimised spectral correlation.

\end{document}